\documentclass[AMA,STIX1COL]{WileyNJD-v2}

\articletype{main paper}%

\received{}
\revised{}
\accepted{}

\usepackage{caption}
\usepackage{subcaption}

\DeclareMathOperator*{\argmin}{argmin}\usepackage{caption}

\begin{document}
\title{Empirical Likelihood Inference for Area under the ROC Curve using Ranked Set Samples}%\protect\thanks{This is an example for title footnote.}}

\author[1]{Chul Moon*}
\author[1]{Xinlei Wang}
\author[2]{Johan Lim}

\authormark{MOON \textsc{et al}}

\address[1]{\orgdiv{Department of Statistical Science}, \orgname{Southern Methodist University, Dallas}, \orgaddress{\state{Texas}, \country{USA}}}

\address[2]{\orgdiv{Department of Statistics}, \orgname{Seoul National University}, \orgaddress{\state{Seoul}, \country{Republic of Korea}}}

\corres{*Chul Moon, Department of Statistical Science, Southern Methodist University, Dallas, Texas, USA. \email{chulm@smu.edu}}

%\presentaddress{This is sample for present address text this is sample for present address text}

\abstract[Summary]{The area under a receiver operating characteristic curve (AUC) is a useful tool to assess the performance of continuous-scale  diagnostic tests on binary classification. In this article, we propose an empirical likelihood (EL) method to construct confidence intervals for the AUC from data collected by ranked set sampling (RSS). The proposed EL-based method enables inferences without assumptions required in existing nonparametric methods and takes advantage of the sampling efficiency of RSS. We show that for both balanced and unbalanced RSS, the EL-based point estimate is the Mann-Whitney statistic, and confidence intervals can be obtained from a scaled chi-square distribution.
Simulation studies and two case studies on diabetes and chronic kidney disease data suggest that using the proposed method and RSS enables more efficient inference on the AUC.
%\textcolor{red}{Simulation studies and two case studies on diabetes and chronic kidney disease data suggest that the proposed method performs better than the existing methods and suggest that using RSS enables more accurate inference on the AUC than simple random sampling.}
}


\jnlcitation{\cname{%
\author{Moon C.}, 
\author{X. Wang}, and 
\author{J. Lim}} (\cyear{}), 
\ctitle{Empirical Likelihood Inference for Area under the ROC Curve using Ranked Set Samples}, \cjournal{Pharm. Stat.}, \cvol{}.}

\maketitle

%\footnotetext{\textbf{Abbreviations:} ROC, receiver operating characteristic; AUC, area under a receiver operating characteristic curve; EL, empirical likelihood; RSS, ranked set sampling; SRS, simiple random sampling; MW, Mann-Whitney; BRSS, balanced ranked set sampling; MELE, maximum empirical likelihood estimator; URSS, unbalanced ranked set sampling; NHANES, The US national health and nutrition examination survey; BMI, body mass index; TotChol, total High-density lipoprotein cholesterol; CKD, chronic kidney disease; GFR, glomerular filtration rate; ACR, albumin to creatinine ratio.}

\section{Introduction}

    A receiver operating characteristic (ROC) curve is used to evaluate the performance of a diagnostic test with binary outcomes.
    Let $X$ and $Y$ denote continuous measurements of non-diseased and diseased subjects with distribution functions $F$ and $G$, respectively.
    The diagnostic test classifies a subject as having the disease if the measurement is greater than the threshold $c$.
    The ROC curve plots the sensitivity (the true positive rate, $p=P(Y \geq c)=1-G(c)$) versus $1-$specificity (the false positive rate, $q=1-P(X \leq c)=F(c)$) of the diagnostic test for all possible threshold values. 
    The ROC curve can be expressed as a function $R(p)=1-G(F^{-1}(p))$, for $0\leq p \leq 1$.
    The area under the ROC curve (AUC), denoted $\delta$, is an effective summary measure of the test's overall accuracy. 
    %A receiver operating characteristic (ROC) curve is used to evaluate the performance of a diagnostic test with binary outcomes.
    %Let $X$ and $Y$ denote continuous measurements of non-diseased and diseased subjects with distribution functions $F$ and $G$, respectively. The diagnosis test classifies a subject as having the disease if the measurement is greater than the threshold $c$.
    %The ROC curve plots the sensitivity (the true positive rate, $p=P(Y \geq c)=1-G(c)$) versus $1-$specificity (the false positive rate, $q=1-P(X \leq c)=F(c)$) of the diagnostic test for all possible threshold values. 
    %The ROC curve can be expressed as a function $R(p)=1-G(F^{-1}(p))$, for $0\leq p \leq 1$.
    %The area under the ROC curve (AUC), denoted $\delta$, is an effective summary measure of the test's overall accuracy. 
    Bamber\cite{Bamber1975} shows that $\delta=P(Y \geq X)$, which can be interpreted as the probability that the measurement of a randomly selected diseased subject is higher than that of a randomly selected non-diseased subject.

Many nonparametric approaches have been widely used to estimate the AUC from simple random samples, including the Mann-Whitney (MW) statistic, the kernel method, and the empirical likelihood (EL) approach. First, the MW statistic is an unbiased estimator of the AUC, and the confidence interval can be obtained using its asymptotic normality \citep{Bamber1975,Hanley1982}. However, the normal approximation often requires a large number of samples or leads to low coverage for high values of the AUC in finite samples \citep{Qin2006}.
Second, the kernel method estimates the continuous ROC curve by replacing the indicator function of the empirical cumulative distribution with kernels\cite{Zou1997,Lloyd1998}. The AUC can be obtained by computing the area under the kernel smoothed ROC curve. 
The kernel method usually requires optimal bandwidth to minimize the estimation error. 
Although optimal bandwidth selection is well studied for the estimation of the density itself, it has not been extensively studied for the estimation of the AUC.
%However, the kernel method has been mainly studied for the estimation of the density itself, and usually requires an optimal bandwidth selection to minimize estimation errors, which may not be well suited for estimating the AUC.
%The kernel method usually requires an optimal bandwidth selection to minimize the estimation error, and has been well studied for the estimation of the density itself. However, the optimal bandwidth may not be well suited for estimating the AUC.
Lastly, the EL approach for the AUC is proposed by Qin and Zhou\cite{Qin2006} to build confidence intervals for the MW estimator. EL is a nonparametric method of inference introduced by Owen\cite{owen1988,Owen1990,owen1991}. EL enables a distribution-free inference while carrying nice properties of the conventional likelihood such as Wilk's theorem \citep{Owen1990} and Bartlett correction \citep{diciccioetal1991}. Regarding the inference on the AUC, 
Qin and Zhou\cite{Qin2006} show that their EL approach provides a better interval estimation in view of the coverage probability and the length of the interval. 	

Ranked set sampling (RSS), firstly proposed by McIntyre\cite{McIntyre1952}, is a cost-effective sampling method compared to simple random sampling (SRS). RSS is especially useful when precise measurement of primary outcomes is expensive but assessment of their relative ranks is feasible. Various topics have been studied on RSS such as distribution functions \citep{Stokes1988}, nonparametric two-sample tests \citep{Bohn1992,Bohn1994}, and rank regression \citep{Ozturk2002}. In the past decade, RSS has been applied to diverse areas including agriculture\citep{Ghosh2017,Ozturk2018,ozturk2019post,Ozturk2019,Hatefi2020}, education\citep{Wang2016}, engineering\cite{Li2019}, environment\citep{Hatefi2015,Frey2017}, public health\citep{Dumbgen2020,Wang2020,Zamanzade2020,faraji2021another,frey2021robust}, and medicine\citep{Omidvar2018,faraji2021another}. %In the past decade, research effort in RSS remains abundant in RSS, e.g., \cite{Frey2019b,Frey2019a,Li2019}. 
%Ozturk2012
%For detailed reviews of RSS, see Chen et.al\cite{Chen2004}, Wolfe\cite{Wolfe2012}, and references therein. 

The main theme of this paper is the inference on the AUC based on EL when data are obtained from RSS. In RSS, the estimation and testing of the AUC have been of great interest. Bohn and Wolfe\cite{Bohn1992,Bohn1994} show the asymptotic normality of the MW statistic of RSS under the location shift distributional assumption $G(x)=F(x-\Delta)$, where $-\infty < \Delta < \infty$. Sengupta and Mukhuti\cite{Sengupta2008} show the efficiency of the MW statistic of RSS over SRS. The kernel methods for RSS also have been introduced \cite{Mahdizadeh2016,Yin2016}. On the other hand, the EL approach for estimating AUC in RSS has not been much studied. Only a few EL methods have been proposed for some specific topics in RSS\citep{Liu2009,Baklizi2009,Baklizi2011}. 
First, Baklizi\cite{Baklizi2009,Baklizi2011} proposes EL methods to construct confidence intervals for population mean and quantiles from one-sample (balanced and unbalanced) RSS data. Thus, these methods cannot be applied to AUC inferences involving two populations.  
	Second, Liu et al.\cite{Liu2009} study EL-based hypothesis testing and interval estimation for one- and two-sample RSS data that mainly focus on mean and mean difference, with extension to general estimating equations. However, their approach is only developed for BRSS. It also does not take full advantage of information obtained from RSS as it assigns equal weight to all measured units in the same cycle of BRSS.
	%Their method assigns equal weight to the measurements in the same cycle without considering individual importance.
	%it fails to fully utilize information obtained from RSS. 
	%it is limited because the sample mean for each cycle are used as an observation in EL, instead of using all RSS samples.
	%For example, for BRSS samples with the single cycle and set size $n$, its sample mean of the cycle $\overline{Y}_{BRSS}$ is used as a single observation in EL. 
	%To overcome these limitations, we propose the EL-based approach for AUC inference that fully utilizes information obtained from samples of BRSS and URSS.
To overcome these limitations, we propose EL-based methods for the AUC interval estimation that may better utilize the information obtained from balanced and unbalanced ranked set samples. Like other EL-based methods, our method does not make any distributional assumptions and does not have an additional parameter to be tuned, such as bandwidth. We illustrate the performance of the proposed EL-method via simulation using purely synthetic data and data re-sampled from two case studies, one involving diabetes data and the other involving chronic kidney disease data. 

The rest of the paper is organized as follows. In Section~\ref{sec:method}, we introduce RSS and basic notations needed for our study, and propose the EL methods to estimate the AUC for balanced and unbalanced RSS. In Section~\ref{sec:sim}, we conduct simulation studies under various settings to compare the performance of the proposed EL-based methods for RSS with the existing EL-based method for SRS and kernel-based method for RSS. 
In Section~\ref{sec:case}, we apply the proposed method to two case studies of diabetes and chronic kidney disease data and compare it with the other known methods. %\textcolor{red}{The simulation and case studies are conducted to show that provided to performance of estimators for much smaller N simulations.}
Discussion and conclusion are given in Section~\ref{sec:conclusion}.

	\section{Methods}
	\label{sec:method}
	\subsection{Ranked Set Sampling}
	Let $Y$ be the variable of interest. The RSS scheme can be described as follows. 
	First, the SRS sample of size $n$ is generated from the population of $Y$. 
	Then the $n$ sample units are ranked by judgment without actually measuring $Y$. If there is an easily available concomitant variable, then it can be used to rank the units approximately.
	%For example, determining whether the hazardous waste site is contaminated ($Y$) can be indirectly examined by the photos and physical characteristics of the site ($C_Y$).
	%determining if a hazardous waste site is polluted ($Y$) could be examined by the photos and physical characteristics of the sites ($C_Y$) \citep{ross1999special}.
	From the $n$ ranked sample units, the $r$th smallest one is chosen for actual quantification and is denoted $Y_{[r]1}$; all the other $n-1$ units are discarded.
	%Here, $n$ number of measured observations are used to obtain one RSS sample.
	Note that this single RSS observation is obtained using auxiliary ranking information from the $n-1$ observations that are only used for ranking.
	This procedure continues until $l_r$ observations $\{ Y_{[r]1},Y_{[r]2},\ldots,Y_{[r]l_r} \}$, all with rank $r$, are obtained and the set of these observations is called the $r$th rank stratum.
	The similar process is repeated for $r= 1,\ldots,n $ so that $n$ strata are obtained. 
	%The total number of ranked set samples of $Y$ is $n_y=\sum\limits_{r=1}^n l_r$. 
	The ranked set sample of $Y$ can be represented as
	%with the in-stratum sample size of $l_r$.
	%s of $(l_1,l_2,\ldots,l_n)$. 
	%As a result, we have $n_y=\sum\limits_{r=1}^n l_r$ total number of ranked set samples of $Y$.
	\begin{eqnarray*}
		    &Y_{[1]1}, Y_{[1]2}, \cdots Y_{[1]l_1}     \nonumber \\
		    &Y_{[2]1}, Y_{[2]2}, \cdots Y_{[2]l_2}     \nonumber \\
		    %&\;\;\; \vdots \;\;\; \ddots \;\;\; \vdots \\
		    &\cdots \;\;\; \cdots \;\;\; \cdots \;\;\; \cdots \\
		    &Y_{[n]1}, Y_{[n]2} \cdots Y_{[n]l_n},
	\end{eqnarray*}
	where the $r$th row is the $r$th rank stratum.
	As a result, it takes $n\sum_{r=1}^nl_r$ number of sample units to obtain $\sum_{r=1}^nl_r$ number of RSS observations. In the above RSS sample, $n$ is referred to as the set size and the total sample size is the total number of units formally quantified, given by $\sum_{r=1}^n l_r$.
	
	In balanced RSS (BRSS), the same number of sample units is assigned to all $n$ strata $(l=l_1=l_2=\ldots=l_n)$, where the $s$th column $\{Y_{[1]s},Y_{[2]s},\ldots,Y_{[n]s} \}$ is called the $s$th cycle with set size $n$.
	%For example, if the samples above are obtained by BRSS, the observations in the $s$th column, $\{Y_{[1]s},Y_{[2]s},\ldots,Y_{[n]s} \}$, are the elements of the $s$th cycle with set size $n$ in the representation above. 
	%It is often said that BRSS samples are obtained from $l$ independent cycles with set size $n$.
	%It is often the case that balanced ranked set samples are obtained by repeating $l$ cycles. 
	On the other hand, unbalanced RSS (URSS) quantifies different number of observations in each rank stratum so that at least one $l_r$ is not equal to the others.

	%optimal set size  Additional information can be obtained by using large set size with perfect judgement ranking. On the other hand, it is possible that large set size may increase ranking error if rankings are not perfect. Therefore, the quality of judgement rankings their impact need to be used to determine the set size.  

	RSS provides more efficient inference compared to SRS. For example, let $\{ Y_{[1]},Y_{[2]},\ldots,Y_{[n]} \}$ be the BRSS samples of size $n$ from a single cycle and $\{Y_1,Y_2,\ldots,Y_n\}$ be the SRS samples of size $n$. 
	The BRSS and SRS estimators for population mean are $\overline{Y}_{BRSS}=\frac{1}{n}\sum_{i=1}^n Y_{[i]}$ and $\overline{Y}_{SRS}=\frac{1}{n}\sum_{i=1}^n Y_i$, respectively. Both estimators are unbiased $E[Y]=E\left[\overline{Y}_{BRSS}\right]=E\left[\overline{Y}_{SRS}\right]$, but the variance of the BRSS estimator is always smaller than or equal to that of the SRS estimator $\text{Var}\left(\overline{Y}_{BRSS}\right)\leq \text{Var}\left(\overline{Y}_{SRS}\right)$ \cite{Chen2004,Wolfe2012}. 
	
	RSS can improve the efficiency of comparisons regardless of the set size. 
	%even if the set size is not optimal.
	The set size and the ranking quality play an important role in determining sampling efficiency in RSS. 
	In practice, small set sizes are preferred because ranking a large number of units by judgment may be difficult and there is no way to guarantee the ranking quality. In our numerical study, we use small set sizes in the set $\{2,4,5\}$.

	\subsection{Empirical Likelihood Method for AUC with Balanced RSS}
	\label{subsec:2.2}

	%\textcolor{red}{We first propose the EL-based method for the AUC using BRSS samples.}
	Let samples of $X$ and $Y$ be obtained by BRSS. Also, let $Y_{[r]s}$ be the observation of a diseased subject in the $r$th sample of the $s$th cycle, which is judged to be the $r$th smallest. We assume that the diseased observations are obtained from $l>1$ cycles, each containing $n$ quantified units with ranks from $1$ to $n$, so that $r=1,\ldots,n$ and $s=1,\ldots,l$. 
	Similarly, the non-diseased observations $X_{[i]j}$'s are obtained by BRSS of $k>1$ cycles with size $m$, $i=1,\ldots,m$ and $j=1,\ldots,k$.

	Let us write 
	\begin{equation*}
	    \phi(X_{[i]j},Y_{[r]s})=\left\{
            \begin{array}{@{}ll@{}}
                    1, & \text{if}\ X_{[i]j}<Y_{[r]s} \\
                    0, & \text{otherwise.}
                    \end{array}\right.
	\end{equation*}
    The MW statistic of the AUC based on BRSS is
		\begin{eqnarray*}
			\widehat{\delta}_{BRSS} &=&  %U_{BRSS}(Y_{[1]1},\ldots,Y_{[n]l};X_{[1]1},\ldots,X_{[m]k}) \\
			U_{BRSS}\left(Y_{[1]1},\ldots,Y_{[1]l},\ldots,Y_{[n]1},\ldots,Y_{[n]l};X_{[1]1},\ldots,X_{[1]k},\ldots,X_{[m]1},\ldots,X_{[m]k}\right) \\
			&=& \frac{1}{mknl} \sum\limits_{i=1}^m \sum\limits_{j=1}^k \sum\limits_{r=1}^n \sum\limits_{s=1}^l  \phi(X_{[i]j},Y_{[r]s}).
		\end{eqnarray*}
	
	%where $\phi(X_{[i]j},Y_{[r]s})=1$  is an indicator function.
	%For convenience, we write %$$ as an indicator function 
	%$\phi(X,Y)=I(X<Y)$ for the rest of the paper.  
	
	We can define the placement value of $Y_{[r]s}$ similar to Pepe and Cai\cite{Pepe2004} as 
	\begin{equation*}
	   U_{rs}=1-F(Y_{[r]s}).
	\end{equation*}
	Here, $U_{rs}$ can be seen as a proportion of non-diseased subjects whose measurements are greater than $Y_{[r]s}$. By treating $Y_{[r]s}$'s as independent and identically distributed (i.i.d.) random variables from the distribution function $G(\cdot)$, the AUC ($\delta$) can be obtained by using $U_{rs}$,
	\begin{equation}
	   E\left[\sum\limits_{r=1}^n\sum\limits_{s=1}^{l} \frac{1}{nl}\left(1-U_{rs}\right)\right]=E[F(Y) ]=\delta.
	   \label{eq:rel}
	\end{equation}
	Here, the stratification is not considered by treating the RSS data as if they were i.i.d. However, this has often been a beneficial approach in BRSS\cite{Baklizi2009}.

	We can derive an EL procedure using the relationship between $U_{rs}$ and $\delta$ given in (\ref{eq:rel}).
	Let $\mathbf{p}=\left( p_{11},\ldots,p_{rs},\ldots,p_{nl} \right)$ be a probability vector such that $\sum_{r=1}^{n}\sum_{s=1}^{l} p_{rs}=1$ and $p_{rs}\geq0$ for all $r=1,\ldots,n$ and $s=1,\ldots,l$. %\textcolor{red}{$p_{rs}=dG(Y_{[r]s})$} for $r=1,\ldots,n$ and $s=1,\ldots,l$.
	The profile EL for the AUC using balanced ranked set samples evaluated at the true AUC value $\delta_0$ is
	\begin{equation}
	L_{BRSS}(\delta_0) = \sup \left\{ \prod\limits_{r=1}^{n}\prod\limits_{s=1}^{l} p_{rs}: \sum\limits_{r=1}^{n}\sum\limits_{s=1}^{l} p_{rs}=1, \sum\limits_{r=1}^n\sum\limits_{s=1}^{l} p_{rs}  \left(1-U_{rs}-\delta_0\right) =0 \right\}.
	\label{eq:1}
	\end{equation}
	%\textcolor{red}{additional explanation?}
	%\begin{equation}
	%   L(\delta_0) = \sup \left\{ \prod\limits_{i=1}^{m} p_{i}: \sum\limits_{i=1}^{m} p_{i}=1, \sum\limits_{i=1}^m p_{i} \left( \sum\limits_{j=1}^k (1-U_{ij}-\delta_0) \right) =0 \right\},
	%\end{equation}
	 
	%We need to estimate $\widehat{U}_{rs}=1-\widehat{F}(Y_{[r]s})$
	Because $U_{rs}$ depends on the unknown non-disease population distribution function $F$, following Qin and Zhou\cite{Qin2006}, we estimate it using the empirical distribution of $\widehat{F}$ as defined in Stokes and Sager\cite{Stokes1988}. %to estimate the non-disease population distribution function $F$.
	By replacing $U_{rs}$ with $\widehat{U}_{rs}=1-\widehat{F}\left(Y_{[r]s}\right)=1-\frac{1}{mk}\sum\limits_{i=1}^m\sum\limits_{j=1}^k\phi\left(X_{[i]j},Y_{[r]s}\right)$ in (\ref{eq:1}) and solving the Lagrange multiplier, we get
	\begin{equation*}
	p_{rs} = \frac{1}{nl}\left( 1+\lambda \left(1-\widehat{U}_{rs}-\delta_0\right) \right)^{-1},
	\end{equation*}
	where $\lambda$ is the solution to
	\begin{equation}
	0 = \frac{1}{nl} \sum\limits_{r=1}^n \sum\limits_{s=1}^l \frac{  \left(1-\widehat{U}_{rs}-\delta_0\right) }{1+\lambda \left(1-\widehat{U}_{rs}-\delta_0\right)}.
	\label{eq:3}
	\end{equation}
	The empirical log-likelihood ratio is
	\begin{equation}
	l_{BRSS}(\delta_0) = 2\sum\limits_{r=1}^n \sum\limits_{s=1}^l \log \left( 1+ \lambda \left(1-\widehat{U}_{rs}-\delta_0\right) \right).
	\label{eq:4}
	\end{equation}

	The maximum empirical likelihood estimator (MELE) $\Tilde{\delta}_{BRSS}$ can be obtained by
	\begin{equation*}
	   \Tilde{\delta}_{BRSS} = \argmin_{\delta} l_{BRSS}(\delta).
	\end{equation*}
	In our setting, $\Tilde{\delta}_{BRSS}=\widehat{\delta}_{BRSS}$ because the dimension of $\delta$ and the constraint is the same and $\widehat{\delta}_{BRSS}$ is the solution of $\sum\limits_{r=1}^n \sum\limits_{s=1}^l \left(1-\widehat{U}_{rs}-\delta\right)=0$ \citep{Qin1994}.
	
	The standard EL theory cannot be applied to equation~(\ref{eq:4}) because $\widehat{U}_{rs}$ are not independent. Instead, we study the limiting distribution of the scaled EL \citep{wang2002b,wang2002a,Wang2004}. The following theorem shows that the asymptotic distribution of $l_{BRSS}(\delta_0)$ follows the scaled chi-square distribution with one degree of freedom. The proof is given in Appendix~\ref{subapp:proof}.
	
	\begin{theorem}
		%Assume that rankings of BRSS are consistent, the true value of the AUC is $\delta_0$, and $E(|F(Y)|^3) < \infty$. For fixed $m$ and $n$ as $k\rightarrow \infty$ and $l \rightarrow \infty$, 
		Assume that rankings of BRSS are consistent and the true value of the AUC is $\delta_0$. For fixed $m$ and $n$ as $k\rightarrow \infty$ and $l \rightarrow \infty$, 
		\begin{equation*}
		r_{BRSS}(\delta_0)l_{BRSS}(\delta_0) \rightarrow \chi^2_1,
		\end{equation*}
		where
		\begin{eqnarray*}
		r_{BRSS}(\delta_0)&=&\frac{mk}{mk+nl} \frac{\sum\limits_{r=1}^n \sum\limits_{s=1}^{l} \frac{1}{nl} \left( 1- \widehat{U}_{rs}-\delta_0 \right)^2}{S_{BRSS}^2},\\
		S_{BRSS}^2&=&\frac{nl \left(S_{BRSS}^{10}\right)^2+ mk \left(S_{BRSS}^{01}\right)^2}{mk+nl},\\ 
		\left(S_{BRSS}^{10}\right)^2 &=& \sum\limits_{i=1}^m \sum\limits_{j=1}^{k} \frac{1}{m(k-1)} \left( V_{BRSS}^{10}(X_{[i]j})- \overline{V}^{10}_{BRSS[i]} \right)^2,\\
		\left(S_{BRSS}^{01}\right)^2 &=& \sum\limits_{r=1}^n \sum\limits_{s=1}^{l} \frac{1}{n(l-1)}  \left( V_{BRSS}^{01}(Y_{[r]s})-\overline{V}^{01}_{BRSS[r]} \right)^2,
		\end{eqnarray*}
		\begin{center}
		    $V_{BRSS}^{10}(X_{[i]j})=\frac{1}{nl}\sum\limits_{r=1}^{n}\sum\limits_{s=1}^{l} \phi(X_{[i]j},Y_{[r]s})$, $V_{BRSS}^{01}(Y_{[r]s})=\frac{1}{mk}\sum\limits_{i=1}^{m}\sum\limits_{j=1}^{k} \phi(X_{[i]j},Y_{[r]s})$,\\ $\overline{V}_{BRSS[i]}^{10}=\frac{1}{k}\sum\limits_{j=1}^k V_{BRSS}^{10}(X_{[i]j})$, and $\overline{V}_{BRSS[r]}^{01}=\frac{1}{l}\sum\limits_{s=1}^l V_{BRSS}^{01}(Y_{[r]s})$.
		\end{center}
		\label{theorem1}
	\end{theorem}

	%The variance of the EL-RSS estimators are smaller than equal to that of EL-SRS estimators of \cite{Qin2006}. Assume that $nl$ diseased and $mk$ non-diseased subjects are sampled using SRS. Then, from equation (8),
	%\begin{eqnarray}
	%\text{Var}(\widehat{\delta}_{\text{EL-RSS}}) &=& \text{Var}(\widehat{\delta}_{\text{EL-SRS}}) + 
	%(\beta-\overline{\beta}) + (\alpha-\overline{\alpha})
	%\left(\frac{1}{n}\sum\limits_{r=1}^n\left(\delta_{[r]}^Y\right)^2 - \delta^2_0 \right)  +  
	%\left( \frac{1}{m}\sum\limits_{i=1}^m \left(\delta_{[i]}^X\right)^2 - \delta^2_0 \right)
	%\end{eqnarray}
	%Also, because $\beta \leq \overline{\beta}$ and $\alpha \leq \overline{\alpha}$,
	%\begin{eqnarray}
	%\text{Var}(\widehat{\delta}_{\text{EL-RSS}}) \leq \text{Var}(\widehat{\delta}_{\text{EL-SRS}}).
	%\end{eqnarray}
	
	%\textcolor{red}{In some observations, RSS . All the sampled X values are between two adjacent rank statistics of Y. In this case, $S^2$ can become zero. We use SRS-EL\ldots}
	
	We can use Theorem~\ref{theorem1} to find an interval estimate for $\delta$. An EL $100(1-\alpha)\%$ confidence interval for the AUC based on BRSS samples can be found as
	\begin{equation}
	CI_{\alpha}(\delta)=\left\{ \delta: r_{BRSS}(\delta)l_{BRSS}(\delta) \leq \chi_{1,1-\alpha}^2 \right\},
	\label{eq:ci}
	\end{equation}
	where $\chi_{1,1-\alpha}^2$ is the $(1-\alpha)$ quanatile of $\chi^2_1$ distribution. By Theorem~\ref{theorem1}, the confidence interval~(\ref{eq:ci}) asymptotically achieves the correct coverage probability $1-\alpha$. For simplicity, we approximate $r_{BRSS}(\delta)$ with $r_{BRSS}(\widehat{\delta}_{BRSS})$ in (\ref{eq:ci}). %In our study, EL ratios and confidence intervals are computed using the R package ``emplik''\cite{Zhou2020}.
	
	\subsection{Empirical Likelihood Method for AUC with Unbalanced RSS}
	\label{sec:urss-el}
	%Unbalanced RSS have different . 
	%\subsection{Equal Set Size}
	
	%\subsection{Unbalanced Ranked Set Sampling}

	We extend the proposed EL method to URSS with equal set sizes. 
	%the number of judgement order statistics could differ. 
	In URSS, the number of observations in each rank stratum is not equal. 
	For the measurements of diseased subjects $Y$, the number of observations of the $r$th rank stratum is $l_r$ for $r=1,\ldots,n$, and the total number of ranked set samples becomes $n_y=\sum\limits_{r=1}^n l_r$.
	%in-stratum sample size of $l_r$
	%In URSS, the $r$th ranked statistic among a set of $n$ diseased subjects, $Y_{[r]}$, is quantified $l_r$ times for $r=1,\ldots,n$. 
	%The number of diseased subjects becomes $n_y=\sum\limits_{r=1}^n l_r$. 
	Similarly, the measurements of non-diseased subjects $X$ are obtained $k_i$ times in the rank $i$ stratum for $i=1,\ldots,m$, and the total number of observations is $n_x=\sum\limits_{i=1}^m k_i$. We assume $l_r>1$ for $r=1,\ldots,n$ and $k_i>1$ for $i=1,\ldots,m$.
	%the $i$th ranked statistic among a set of $m$ non-diseased subjects $X_{[i]}$ is obtained $k_i$ times for $i=1,\ldots,m$ and the total number of subjects is $n_x=\sum\limits_{i=1}^m k_i$.
	
	%\begin{align*}
	%	    &Y_{[1]1} \cdots Y_{[1]l_1}           && X_{[1]1} \cdots X_{[1]k_1}\nonumber \\
	%	    &\;\;\; \vdots \;\;\; \ddots \;\;\; \vdots &&\;\;\; \vdots \;\;\; \ddots \;\;\; \vdots \\
	%	    &Y_{[n]1} \cdots Y_{[n]l_n}           && X_{[m]1} \cdots X_{[m]k_m}
	%	\end{align*}
		
	The MW statistic of the AUC based on URSS is 
		\begin{eqnarray*}
			\widehat{\delta}_{URSS}
			&=&  %U_{BRSS}(Y_{[1]1},\ldots,Y_{[n]l};X_{[1]1},\ldots,X_{[m]k}) \\
			U_{URSS}\left(Y_{[1]1},\ldots,Y_{[1]l_1},\ldots,Y_{[n]1},\ldots,Y_{[n]l_n};X_{[1]1},\ldots,X_{[1]k_1},\ldots,X_{[m]1},\ldots,X_{[m]k_m}\right) \\
			&=& \sum\limits_{i=1}^m \sum\limits_{j=1}^{k_i} \sum\limits_{r=1}^n  \sum\limits_{s=1}^{l_r}  \frac{1}{mk_i n l_r} \phi(X_{[i]j},Y_{[r]s}).
		\end{eqnarray*}
	%Using the relationship between $U_{rs}$ and $\delta$,  $E\left[\sum\limits_{r=1}^n\sum\limits_{s=1}^{l_r} \frac{1}{nl_r}\left(1-U_{rs}\right)\right]=\delta$, the profile EL for the AUC of URSS evaluated at the true value $\delta_0$ can be defined as
	
	Using the relationship between $U_{rs}$ and the AUC $\delta$ and treating the data as i.i.d., we can show that $E\left[\sum\limits_{r=1}^n\sum\limits_{s=1}^{l_r} \frac{1}{nl_r}\left(1-U_{rs}\right)\right]=\delta$. Let $\mathbf{w}=\left( w_{11},\ldots,w_{1l_1},\ldots,w_{rs},\ldots,w_{nl_n} \right)$ be a weight vector such that $\sum_{r=1}^{n}\sum_{s=1}^{l} \frac{w_{rs}}{n l_r}=1$ where $w_{rs}\geq0$ for all $r=1,\ldots,n$ and $s=1,\ldots,\max\{l_1,\ldots,l_n\}$. The profile EL for the AUC of URSS evaluated at the true value $\delta_0$ can be defined as

	\begin{equation}
	L_{URSS}(\delta_0) = \sup \left\{ \prod\limits_{r=1}^{n}\prod\limits_{s=1}^{l_r} w_{rs}: \sum\limits_{r=1}^{n}\sum\limits_{s=1}^{l_r} \frac{w_{rs}}{n l_r}=1, \sum\limits_{r=1}^n \sum\limits_{s=1}^{l_r} \frac{w_{rs}}{n l_r} \left(1-U_{rs}-\delta_0\right) =0 \right\}.
	\label{eq:3-1}
	\end{equation}
	The first constraint in equation~(\ref{eq:3-1}) implies that $w_{rs}$ is weighted by the number of measurements in the $r$th rank stratum $l_r$.
	By replacing $U_{rs}$ with $\widehat{U}_{rs}=1-\widehat{F}(Y_{[r]s})= 1-\sum\limits_{i=1}^m \sum\limits_{j=1}^{k_i} \frac{1}{mk_i} \phi(X_{[i]j},Y_{[r]s})$ and solving (\ref{eq:3-1}),
	\begin{equation*}
	w_{rs} = \frac{1}{n_y}\left( 1+\lambda \frac{1}{nl_r} \left(1-\widehat{U}_{rs}-\delta_0\right) \right)^{-1}.
	\end{equation*}
	
	The empirical log-likelihood ratio becomes
	\begin{equation}
	l_{URSS}(\delta_0) = 2\sum\limits_{r=1}^n \sum\limits_{s=1}^{l_r} \log \left( 1+ \lambda \frac{1}{nl_r}\left( 1-\widehat{U}_{rs}-\delta_0 \right) \right).
	\label{eq:urss}
	\end{equation}
	It can be easily shown that the MELE of (\ref{eq:urss}) $\Tilde{\delta}_{URSS}=\widehat{\delta}_{URSS}$ for the same reason explained in Section~\ref{subsec:2.2}.
	
	The proposed EL-based method for BRSS can be easily extended to URSS. Theorem~\ref{theorem2} shows the asymptotic result of the empirical log-likelihood ratio for URSS. The proof is given in Appendix~\ref{subapp:proof}. We can obtain an $100(1-\alpha)\%$ EL confidence interval for the AUC with URSS by %using equation~(\ref{eq:ci}).
	\begin{equation}
	CI_{\alpha}(\delta)=\left\{ \delta: r_{URSS}(\delta)l_{URSS}(\delta) \leq \chi_{1,1-\alpha}^2 \right\}.
	\label{eq:ci2}
	\end{equation}
	Similarly, we use $r_{URSS}(\widehat{\delta}_{URSS})$ to approximate $r_{URSS}(\delta)$ in (\ref{eq:ci2}).
	
	\begin{theorem}
		%Assume that rankings of URSS are consistent, the true value of the AUC is $\delta_0$, and $E(|F(Y)|^3) < \infty$. For fixed $m$ and $n$ as  $\min\limits_r l_r \rightarrow \infty$ and $\min\limits_i k_i \rightarrow \infty$, 
		Assume that rankings of URSS are consistent and the true value of the AUC is $\delta_0$. For fixed $m$ and $n$, as  $\min\limits_r l_r \rightarrow \infty$ and $\min\limits_i k_i \rightarrow \infty$, 
		\begin{equation*}
		r_{URSS}(\delta_0)l_{URSS}(\delta_0) \rightarrow \chi^2_1,
		\end{equation*}
		where
		\begin{eqnarray*}
		r_{URSS}(\delta_0)&=&\frac{n_x}{n_y+n_x} \frac{\sum\limits_{r=1}^n \sum\limits_{s=1}^{l_r} \frac{1}{nl_r} \left( 1- \widehat{U}_{rs}-\delta_0 \right)^2}{S_{URSS}^2},\\
		S_{URSS}^2&=&\frac{ n_y \left(S_{URSS}^{10}\right)^2+ n_x \left(S_{URSS}^{01}\right)^2}{n_x+n_y} ,\\ 
		\left(S_{URSS}^{10}\right)^2 &=& \sum\limits_{i=1}^m \sum\limits_{j=1}^{k_i} \frac{1}{m(k_i-1)} \left( V_{URSS}^{10}(X_{[i]j})- \overline{V}_{URSS[i]}^{10} \right)^2,\\
		\left(S_{URSS}^{01}\right)^2 &=& \sum\limits_{r=1}^n \sum\limits_{s=1}^{l_r} \frac{1}{n(l_r-1)}  \left( V_{URSS}^{01}(Y_{[r]s})-\overline{V}_{URSS[r]}^{01} \right)^2,
		\end{eqnarray*}
		\begin{center}
		    $V_{URSS}^{10}(X_{[i]j})=\sum\limits_{r=1}^{n}\sum\limits_{s=1}^{l_r} \frac{1}{nl_r}\phi(X_{[i]j},Y_{[r]s})$, $V_{URSS}^{01}(Y_{[r]s})=\sum\limits_{i=1}^{m}\sum\limits_{j=1}^{k_i} \frac{1}{mk_i}\phi(X_{[i]j},Y_{[r]s})$,\\ $\overline{V}_{URSS[i]}^{10}=\frac{1}{k_i}\sum\limits_{j=1}^{k_i} V_{URSS}^{10}(X_{[i]j})$, and $\overline{V}_{URSS[r]}^{01}=\frac{1}{l_r}\sum\limits_{s=1}^{l_r} V_{URSS}^{01}(Y_{[r]s})$.
		\end{center}
		\label{theorem2}
	\end{theorem}

	\section{Simulation Studies}
	\label{sec:sim}
	\subsection{Balanced RSS}
	\label{subsec:4-1}
	In this section, we compare simulation results of our proposed EL intervals of BRSS (BRSS-EL) to the two existing methods mentioned in the introduction: EL intervals of simple random sampling (SRS-EL) of Qin and Zhou\cite{Qin2006} and kernel-estimation intervals of BRSS (BRSS-KER) of Yin et al.\cite{Yin2016}. We use SRS-EL as a reference simple random sampling method because it outperforms the other SRS-based methods for inference of the AUC \citep{Qin2006}. For BRSS-KER, the Gaussian kernel is used and the bandwidths are selected as $h_x=0.9\min(s_x, iqr_x/1.34)n_x^{-0.2}$ and $h_y=0.9\min(s_y, iqr_y/1.34)n_y^{-0.2}$, where $s$ is the standard deviation and $iqr$ is the interquartile range, suggested by Silverman\cite{Silverman1986}.
	We evaluate performances of these methods using coverage probabilities and average lengths of confidence intervals. 
	%The relative efficiencies are defined by $MSE(SRS)/MSE(Methods)$.
	
	We sample the measurements of non-diseased subjects $X$ and diseased subjects $Y$ from three sets of distributions: normal, log-normal, and uniform. In the first simulation, $X$ follows the standard normal $N\left(\mu=0,\sigma^2=1\right)$, and $Y$ follows $N\left(\sqrt{5} \Phi^{-1}(\delta) ,4\right)$, where $\Phi$ is cdf of the standard normal distribution. 
	The second simulation generates samples of $X$ from the standard log-normal $LN\left(\mu=0,\sigma^2=1\right)$ and $Y$ from $LN\left(\sqrt{5}\Phi^{-1}(\delta),4\right)$. 
	In the third simulation, we sample $X$ from the standard uniform $U(0,1)$ and $Y$ from $U\left(0, \left(2(1-\delta)\right)^{-1}\right)$.
	%, gamma $\Gamma(5,1)$, and $t_3$ distribution of degrees of freedom 3 
	%$\Gamma(\theta,0.2)$, and $t_3 - \theta$. 
	In all simulation studies, we use four AUC values $\delta \in \left\{0.6,0.8,0.9,0.95 \right\}$.

	For each simulation setting, we generate BRSS samples 5,000 times with sample sizes $n_x=n_y \in \left\{20,40,80 \right\}$ from non-diseased and diseased groups. For BRSS, we use the set sizes $m=n\in\left\{2,4\right\}$ and the number of cycles are determined by $k=n_x$/$m$ and $l=n_y$/$n$. 
	The judgment ranking is done by concomitant variables
	\begin{eqnarray*}
		C_X &=& \rho_X \left( \frac{X-\mu_X}{\sigma_X} \right) + (1-\rho_X^2)Z_X \\
		C_Y &=& \rho_Y \left( \frac{Y-\mu_Y}{\sigma_Y} \right) + (1-\rho_Y^2)Z_Y,
	\end{eqnarray*}
	where ($\mu_X$, $\sigma_X$) and ($\mu_Y$, $\sigma_Y$) are the means and standard deviations of $X$ and $Y$, $Z_X$ and $Z_Y$ follow the independent standard normal distribution, and $\rho_X$ and $\rho_Y$ are the Pearson correlations that control the quality of judgment rankings. We set $\rho_X=\rho_Y \in \left\{0.7, 0.9, 1\right\}$ that represent poor, good, and perfect judgment rankings, respectively.

	\begin{figure}[!ht]
        \centerline{\includegraphics[width=0.65\textwidth]{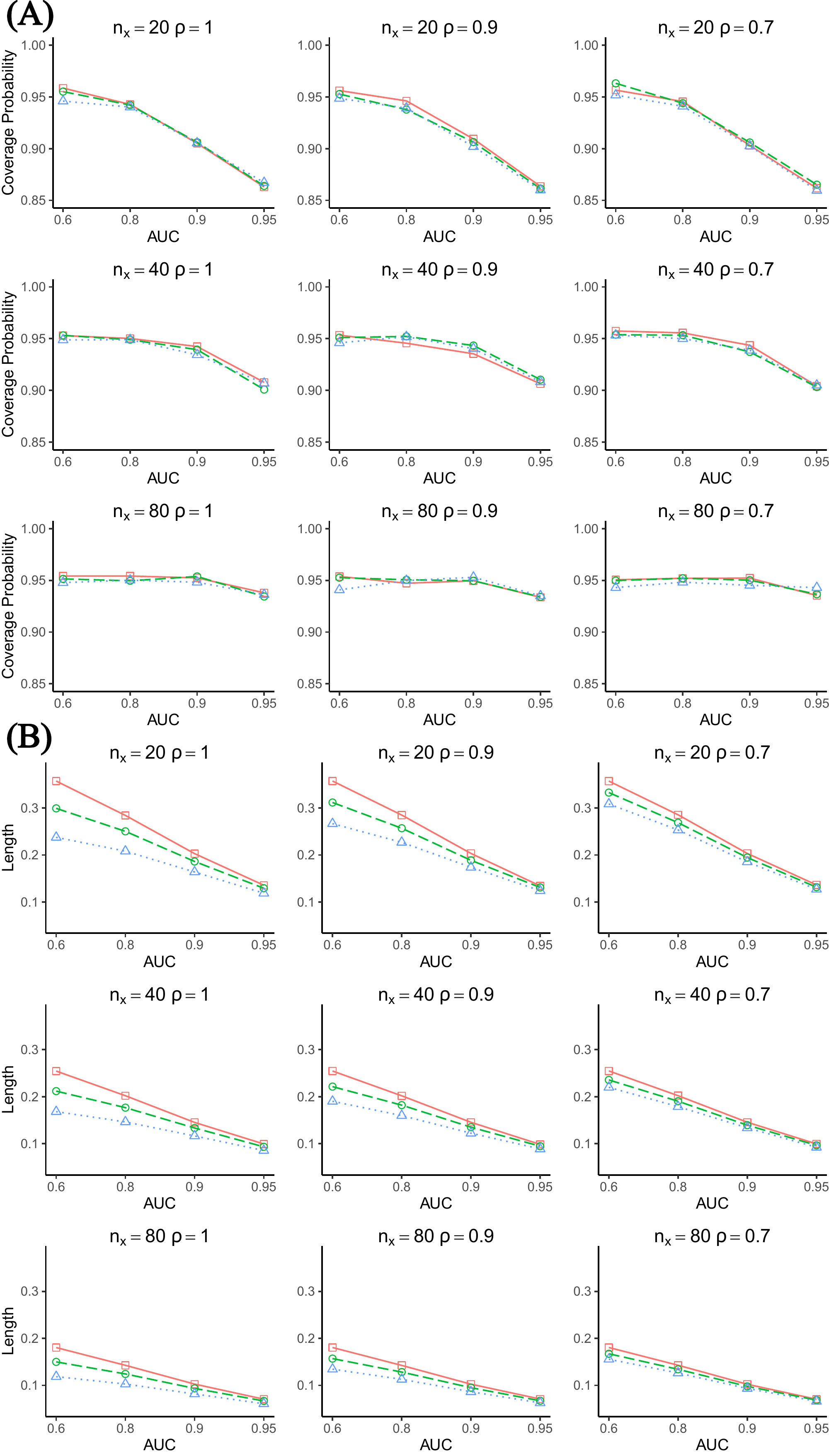}}
	    \caption{Estimated coverage probabilities and average lengths for 95\% confidence intervals of the AUC for the normal distributions: (A) coverage probabilities, (B) average lengths. Three lines are shown: SRS-EL as solid lines with square markers ($\square$), BRSS-EL with $m=2$ as long-dashed lines with circle markers ($\circ$), and BRSS-EL with $m=4$ as dotted lines with triangle markers ($\triangle$).}
	    \label{fig:normal.cp}
	\end{figure}
	
	\begin{figure}[!ht]
	    \centerline{\includegraphics[width=0.65\textwidth]{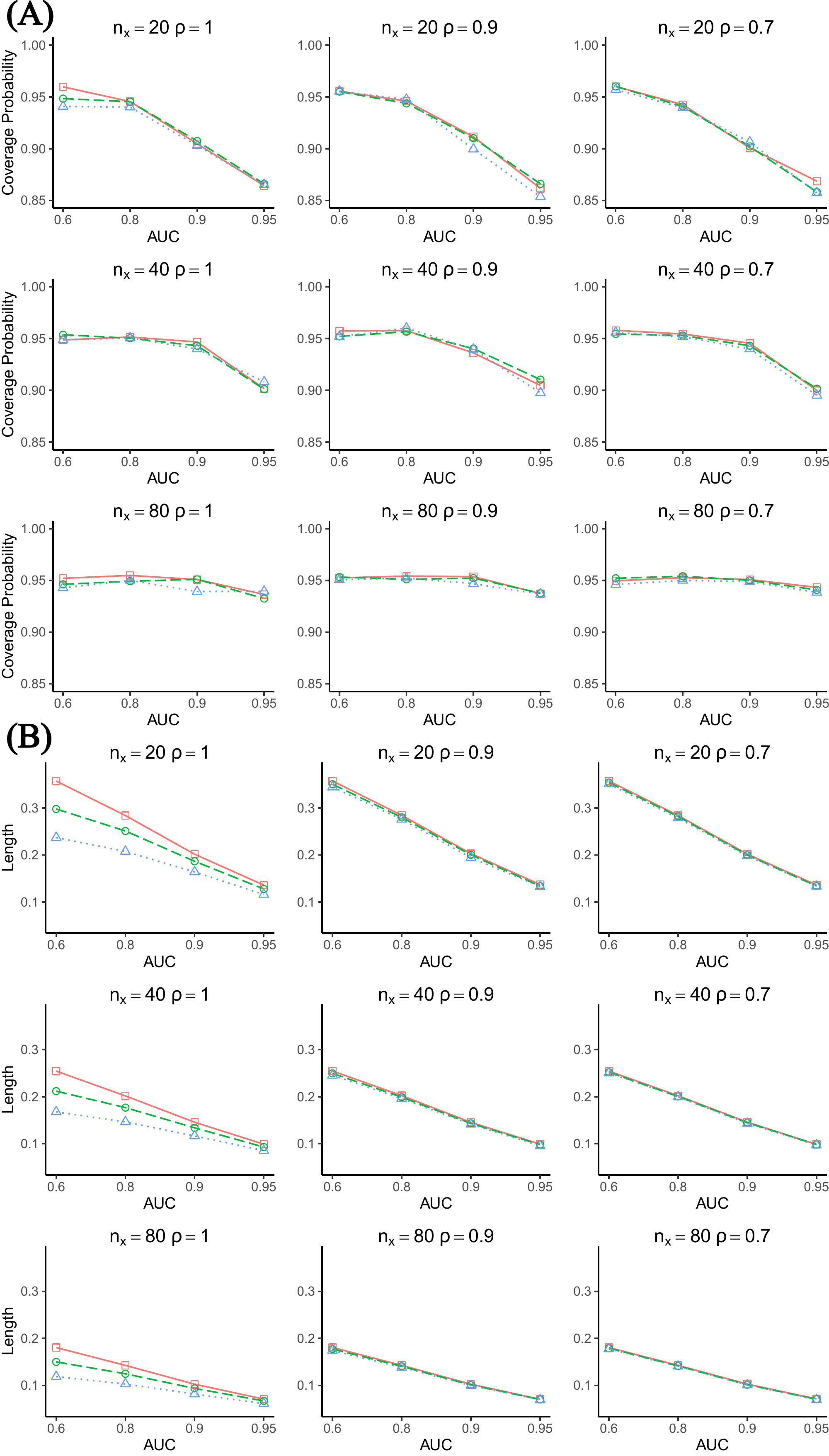}}
	    \caption{Estimated coverage probabilities and average lengths for 95\% confidence intervals of the AUC for the log-normal distributions: (A) coverage probabilities, (B) average lengths. Three lines are shown: SRS-EL as solid lines with square markers ($\square$), BRSS-EL with $m=2$ as long-dashed lines with circle markers ($\circ$), and BRSS-EL with $m=4$ as dotted lines with triangle markers ($\triangle$).}
	    \label{fig:lognormal.cp}
	\end{figure}
	
	\begin{figure}[!ht]
	    \centerline{\includegraphics[width=0.65\textwidth]{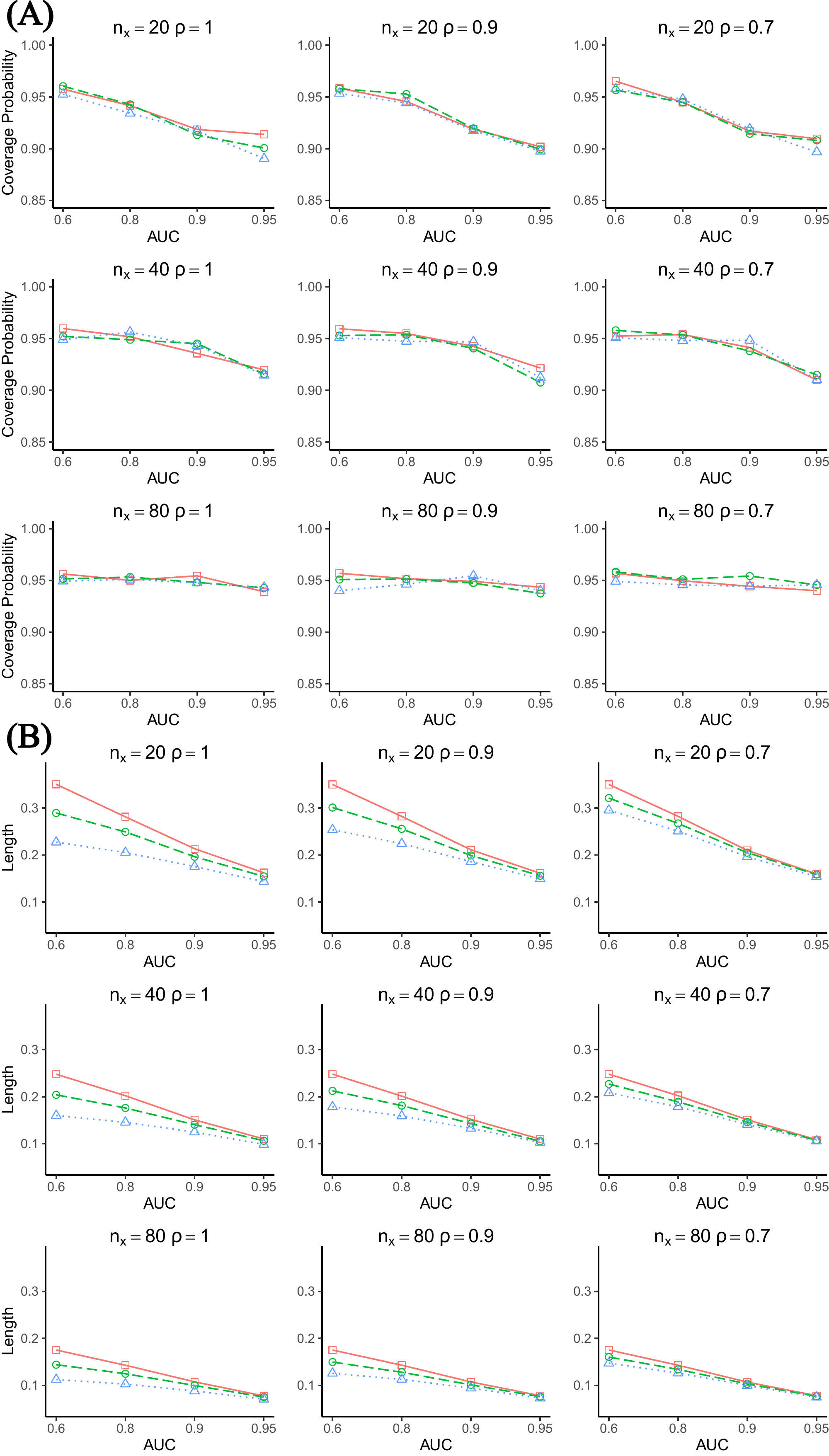}}
	    \caption{Estimated coverage probabilities and average lengths for 95\% confidence intervals of the AUC for the uniform distributions: (A) coverage probabilities, (B) average lengths. Three lines are shown: SRS-EL as solid lines with square markers ($\square$), BRSS-EL with $m=2$ as long-dashed lines with circle markers ($\circ$), and BRSS-EL with $m=4$ as dotted lines with triangle markers ($\triangle$).}
	    \label{fig:uniform.cp}
	\end{figure}
	
	 %The full simulation results are shown in Tables~\ref{tb:normal}, \ref{tb:logn}, and \ref{tb:uniform} in Supplementary Material.
	
	The simulation results indicate that the proposed BRSS-EL outperforms the other methods. Figures~\ref{fig:normal.cp}, \ref{fig:lognormal.cp}, and \ref{fig:uniform.cp} present the coverage probabilities and the average lengths for 95\% confidence intervals of three methods: SRS-EL, BRSS-EL with $m=2$, and BRSS-EL with $m=4$. 
	Except for small sample sizes and high AUC cases where efficient AUC estimation is difficult, the coverage probabilities of all three methods are close to the nominal level.
	BRSS-EL achieves shorter intervals in most cases while achieving similar coverage compared to SRS-EL. 
	
	The quality of judgment ranking affects the average length of RSS intervals.
	As $\rho$ gets smaller so that the quality of the judgment gets poorer, the BRSS-EL intervals get wider and become closer to SRS-EL.
	%However, the coverage probabilities of proposed methods do not change.
	However, even with poor judgement ranking quality with $\rho=0.7$, the proposed method has a shorter interval than SRS-EL.
	
	The EL-based methods show higher coverage than BRSS-KER, particularly when the AUC is close to one.
	The results of BRSS-KER are reported in Figures~\ref{fig:normal.cp.full}, \ref{fig:lognormal.cp.full}, and \ref{fig:uniform.cp.full} in Appendix~\ref{subapp:tbfig} because their coverage probabilities are low compared to the EL-based methods. Especially when the observations are sampled from the log-normal distributions with large AUC values, BRSS-KER performs poorly. The kernel estimates of the AUC are under-estimated, so BRSS-KER intervals usually do not include the true AUC value.
	%The complete results are given .
	
	%Figures~\ref{fig:normal.cp}, \ref{fig:lognormal.cp}, and \ref{fig:uniform.cp} present coverage probabilities and average lengths for 95\% confidence intervals. 
	%For simplicity, results of five methods are shown: SRS-EL, BRSS-EL with $m=2$, BRSS-EL with $m=4$, BRSS-KER with $m=2$, and BRSS-KER with $m=4$.
	%BRSS-EL has shorter intervals while achieving similar coverage accuracy compared to SRS-EL. 
	%Also, the EL-based methods show higher coverage accuracy than BRSS-KER, particularly when the AUC is close to one.
	%Although BRSS-KER has the shortest confidence intervals, their coverage probabilities are less than the EL-based methods in most cases. 
	%Especially when the observations are sampled from the log-normal distributions with large AUC values, BRSS-KER performs poorly. The kernel estimates of the AUC are under-estimated, so the intervals usually do not include the true AUC value.

	%under-estimate the AUC. 
	%The proposed methods achieve higher coverage probabilities with shorter confidence intervals than other methods. 
	%SRS-EL have similar coverage probability as our proposed method. However, the lengths are longer.  
	%The kernel-based methods have low coverage probabilities when the sample size and AUC are large. 

	%Coverage probabilities of large AUC are small when the sample size is small

	We also compare the computation time of BRSS-EL, SRS-EL, and BRSS-KER. We measure the time to run the simulations under three set sizes $m=n\in\{2,4,5\}$ with sample sizes $n_x=n_y=20$, $\rho=1$, $\delta=0.6$, and $X \sim N\left(0,1\right)$, and $Y\sim N\left(\sqrt{5} \Phi^{-1}(0.6) ,4\right)$. The SRS and BRSS samples are generated 5,000 times for each simulation setting and Intel Xeon E5-2695v4 2.1 GHz processors are used for computations. The total execution time for 5,000 replicates in each setting is reported in Table~\ref{table:time} in Appendix~\ref{subapp:tbfig}. The EL-based methods (BRSS-EL and SRS-EL) take longer time than the kernel-based method (BRSS-KER) due to the computation of EL. The computation time of BRSS-EL is similar to that of SRS-EL, but it tends to decrease as the set size increases with fixed sample sizes. 
	
	\subsection{Unbalanced RSS}
	\label{subsec:3.2}
	
	We compare the EL-based methods using samples obtained by URSS (URSS-EL) and SRS (SRS-EL). For simplicity, only $Y$ is sampled by URSS whereas $X$ is obtained by BRSS with set sizes $m=n=2$. The number of the first and the second ranked sets of $Y$ is determined by the proportion $p_y$ such that $l_1=n_yp_y$ and $l_2=n_y(1-p_y)$. When $p_y=0.5$, $Y$ is sampled by BRSS. We set $p_y \in \{ 0.3,0.4,0.5,0.6,0.7\}$ and assume perfect judgement ranking $\rho_X=\rho_Y=1$.
	The other simulation settings are the same as Section~\ref{subsec:4-1}: URSS samples are generated 5,000 times for two sub-populations, $X$ and $Y$ are sampled from the three sets of distributions (normal, log-normal, and uniform), four AUC values are used $\delta \in \{ 0.6,0.8,0.9,0.95\}$, and sample sizes are $n_x=n_y \in \left\{20,40,80 \right\}$. We also randomly generate SRS samples with the same AUC and sample size settings and compute SRS-EL results. The simulation results of BRSS-KER are not reported because it is proposed only for BRSS.
	
	\begin{figure}[!ht]
	    \centerline{\includegraphics[width=0.88\textwidth]{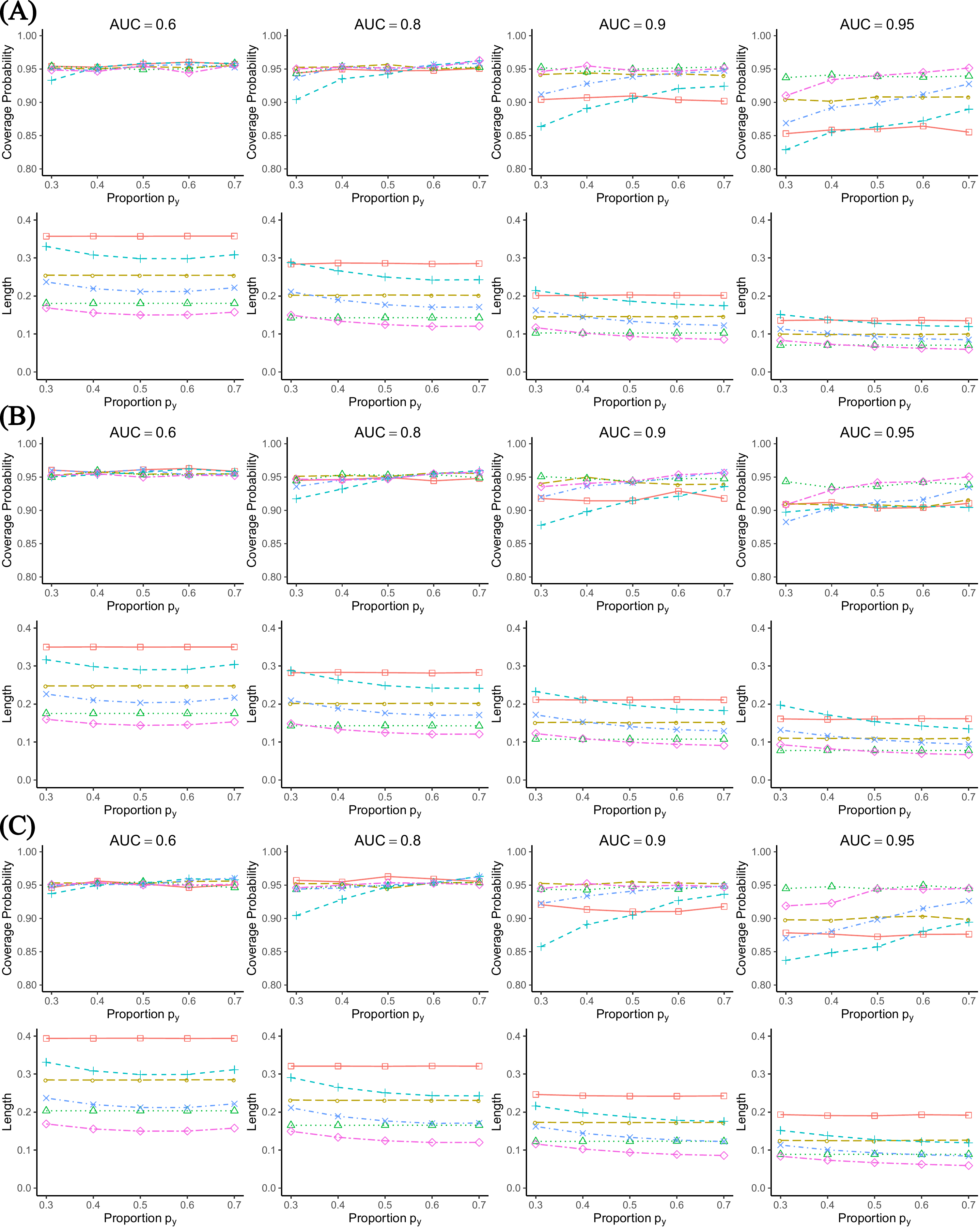}}
	    \caption{Coverage probabilities and average lengths for 95\% confidence intervals of the AUC: (A) normal distribution, (B) log-normal distribution, (C) uniform distribution. Each plot presents six lines: SRS-EL of $n_y=20$ as solid lines with square markers ($\square$), URSS-EL of $n_y=20$ as dashed lines with cross markers ($+$), SRS-EL of $n_y=40$ as long-dashed lines with circle markers ($\circ$), URSS-EL of $n_y=40$ as dot-dashed lines with times markers ($\times$), SRS-EL of $n_y=80$ as dotted lines with triangle markers ($\triangle$), and URSS-EL of $n_y=80$ as two-dashed lines with diamond markers ($\diamond$).}
	    \label{fig:urss}
	\end{figure}

	URSS-EL performs better than BRSS-EL and SRS-EL under the well-designed URSS scheme.
	Figure~\ref{fig:urss} shows results of URSS-EL and SRS-EL simulations.
	%The proportion $p_y$ affects the performance of URSS-EL.
	%When $p_y$ is small, URSS-EL yields lower coverage rates and longer intervals than SRS-EL of the same sample size in some settings.
	%When $p_y$ is small(large), URSS-EL yields lower(higher) coverage rates and longer(shorter) intervals than SRS-EL of the same sample size \textcolor{red}{in some settings}.
	%As shown in the previous simulations, similar when BRSS.
	%The URSS-EL intervals are shorter than the SRS-EL intervals the of the same sample size in most settings. 
	%The coverage rates of URSS-EL vary the proportion $p_y$. When $p_y$ is small, it could be worse than SRS-EL. 
	When the AUC is close to one, taking a larger number of the first-ranked set samples of $Y$ (i.e., larger $p_y$ values) leads to coverage probabilities close to the nominal level and shorter intervals. 
	In this case, the first-ranked set samples of $Y$ provide more information than the second-ranked set samples for the AUC estimation.
	Under the high AUC, the second-ranked set sample of $Y$ is most likely to be greater than the sample of $X$, so it is less informative.
	Therefore, obtaining a larger number of the first-ranked set samples of $Y$ enables better comparison of $F$ and $G$ for the high AUC.
	On the other hand, when the AUC is relatively low, the shortest interval of URSS-EL is achieved when samples are obtained by BRSS.
	
	\section{Case Studies}
	\label{sec:case}
	\subsection{Application to Diabetes Data}
	\label{subsec:5.1}
	We compare the AUC estimation methods using the US National Health and Nutrition Examination Survey (NHANES) collected from 2009 to 2010, available in the R package ``NHANES'' \citep{NHANES}. 
	The NHANES dataset includes demographic, physical, and health information of 10,000 individuals. 
	Although the NHANES data were obtained under complex survey design, we only consider two sub-populations: subjects who do not have diabetes ($X$) and have diabetes ($Y$). One of the best predictors for diabetes available in the NHANES dataset is the body mass index (BMI). The AUC value estimated using BMI is 0.73. 
	
	%We generate samples of size $n_x=n_y \in \left\{20,40,60,80 \right\}$ from two groups, with and without diabetes. For BRSS, two set sizes are used $m=n\in\left\{2,4\right\}$ and the numbers of cycles are determined by $k=n_x$/$m$ and $l=n_y$/$n$. In each simulation setting, we repeat the sampling process 5,000 times.
	To generate SRS/BRSS samples, we treat the entire dataset of the 10,000 individuals as the ``true'' population, which is divided into two sub-populations, individuals with and without diabetes. In this simulation, we first set $n_x=n_y$  and let the sample sizes vary in the set $\{20, 40, 60, 80\}$. For BRSS samples, we set $m=n$ and let the set sizes vary in the set $\{2,4\}$, and the numbers of cycles are determined by $k=n_x$/$m$ and $l=n_y$/$n$. In each simulation setting, we repeat the sampling procedure 5,000 times.
	
	We use two concomitant variables, weight and total High-density lipoprotein cholesterol (TotChol), to illustrate how the results are affected by the quality of ranking. The weight has high Pearson correlations with BMI ($\rho_X=0.897$ and $\rho_Y=0.881$) thus it makes the high quality rankings.
	One the other hand, the TotChol has small negative Pearson correlations with BMI ($\rho_X=-0.294$ and $\rho_Y=-0.274$), which makes the poor quality rankings.
	%The ranking procedure is done by two concomitant variables, weight and total HDL cholesterol (TotChol). The weight has high Pearson correlations with BMI ($\rho_x=0.897$ and $\rho_y=0.881$) whereas the TotChol has low Pearson correlations ($\rho_x=-0.294$ and $\rho_y=-0.274$). 
	
	\begin{figure}[!ht]
	    \centerline{\includegraphics[width=0.5\textwidth]{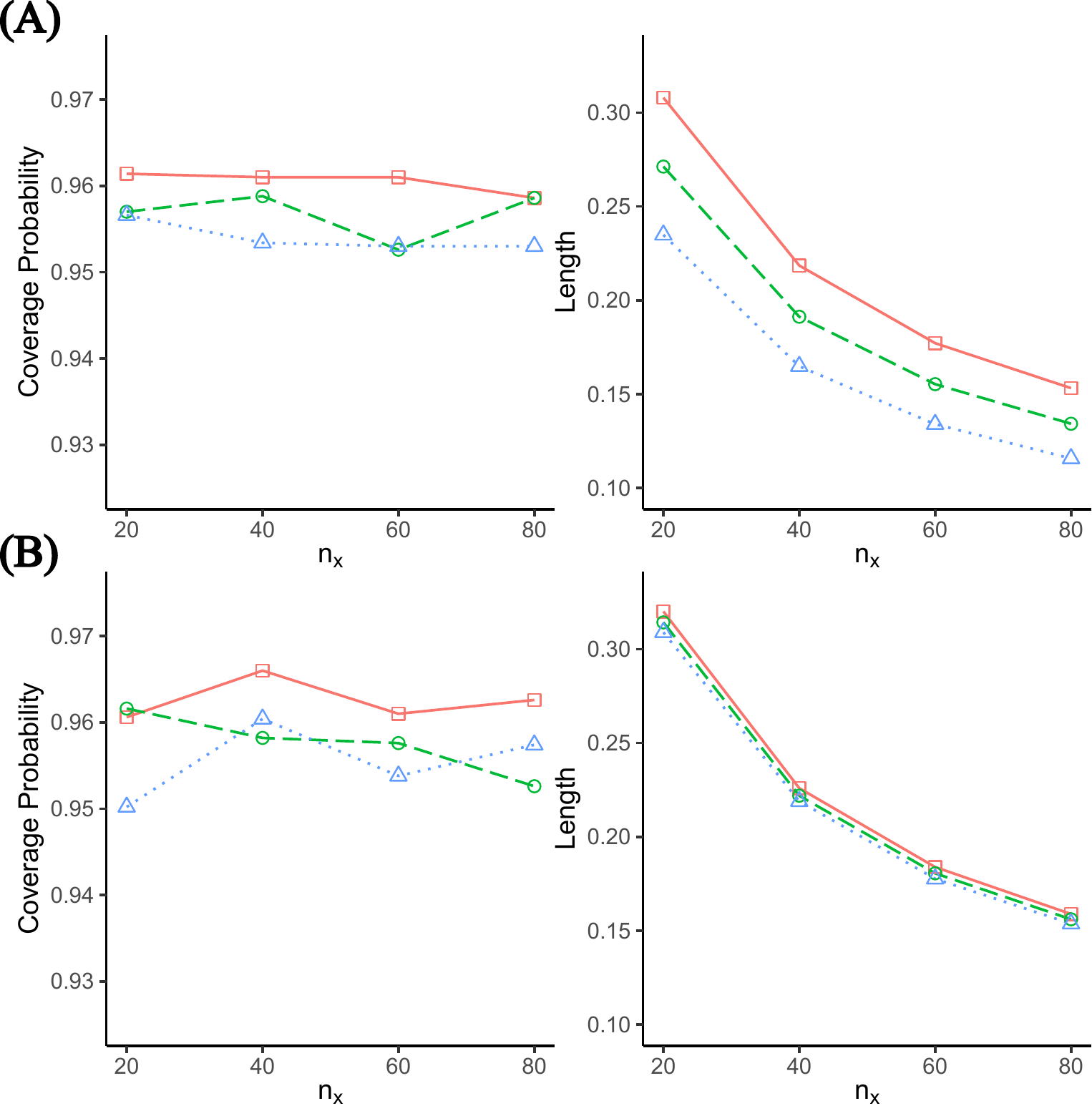}}
	    \caption{Coverage probabilities and average lengths for 95\% confidence intervals of the AUC for the diabetes data: (A) weight is used as concomitant variable, (B) TotChol is used as concomitant variable. Three lines are shown: SRS-EL as solid lines with square markers ($\square$), BRSS-EL with $m=2$ as long-dashed lines with circle markers ($\circ$), and BRSS-EL with $m=4$ as dotted lines with triangle markers ($\triangle$).}
	    \label{fig:diabetes}
	\end{figure}

	Figure~\ref{fig:diabetes} shows coverage probabilities and average lengths for 95\% confidence intervals of SRS-EL, BRSS-EL with $m=2$, and BRSS-EL with $m=4$.
	When the weight is used for judgement ranking, BRSS-EL has shorter intervals than SRS-EL while maintaining similar coverage probabilities.
	On the other hand, when the TotChol is used as the concomitant variable, the BRSS-EL intervals are still slightly shorter than the SRS-EL intervals despite the lower ranking quality. In both cases, coverage probabilities of BRSS-EL are closer to 0.95 than that of SRS-EL.
	The complete results that include BRSS-KER are given in Figure~\ref{fig:diabetes.full} in Appendix~\ref{subapp:tbfig}. BRSS-KER has the shortest intervals, but its coverage probabilities are the lowest. 

	%Figure~\ref{fig:diabetes} shows coverage probabilities and average lengths for 95\% confidence intervals.
	%When the weight is used for ranking, BRSS-EL has shorter intervals than SRS-EL while maintaining similar coverage probabilities. BRSS-KER has the shortest intervals, but its coverage probabilities are the lowest. 
	%On the other hand, when the TotChol is used as the concomitant variable, the BRSS-EL intervals are still slightly shorter than SRS-EL though the ranking quality is low.
	
	The result suggests that BRSS-EL yields better inferences over SRS-EL regardless of the quality of the rankings. If the concomitant variable is informative, such as the weight in this application, the proposed method gives more efficient inference than the other methods. 
	Even in the worst case when the concomitant variable is not informative, such as the TotChol, BRSS-EL achieves results similar to SRS-EL. As a result, our method gives better or at least as good results as SRS-EL.
	
	\subsection{Application to Chronic Kidney Disease Data}
	%A person with diabetes can easily develop kidney disease and 
	Chronic kidney disease (CKD) refers to a condition that the kidneys are gradually damaged and cannot filter blood as needed.
	Diabetes is a major cause of CKD. In 2008, about 44\% of new kidney failure was due to diabetes \citep{centers2011national}. 
	An early diagnosis of CKD is important because early treatment may prevent the kidney from being further damaged. Therefore, it is suggested that a person with diabetes monitors the signs of CKD.
	
	There are two common markers that evaluate kidney functions: 1) a blood test that estimates glomerular filtration rate (GFR), and 2) a urine test that measures albumin to creatinine ratio (ACR). 
	GFR estimates how much blood is flowing each minute and it indicates how well a kidney is working. GFR is estimated using multiple factors such as serum creatinine level, age, gender, and ethnicity.
	ACR is used to estimate the excretion of urinary albumin. The damaged kidney does not filter proteins well, and it leads to a high level of urinary albumin, one of the proteins that can be found in urine.
	If people have GFR less than 60 $ml/min/1.73m^2$ or ACR greater than 30 $mg/g$ for three months, then they are at risk of decreased kidney function \citep{levin2013kidney}.
	
	We use 3,051 subjects with diabetes ages 21 to 79 in the NHANES datasets from 2009 to 2018.
	We consider two sub-populations that do not have CKD ($X$) and has CKD ($Y$). The subjects are classified to have CKD if their GFR$<$60 or ACR$>$30. We want to note that three months of GFR and ACR information is needed to diagnose CKD. However, we only use one-time GFR and ACR records for simplicity.
	We estimate GFR using the chronic kidney disease epidemiology collaboration equation \citep{levey2009new}, which is recommended for reporting GFR in adults \citep{earley2012estimating}.
	%as recommended by the National Kidney Foundation.

	We estimate the AUC for CKD among subjects with diabetes using GFR.
	%\textcolor{blue}{GFR is one of the major factors to determine CKD} 
	The estimated AUC calculated by the negative GFR is $0.757$. The negative GFR is used here because low GFR indicates poor kidney function. 
	One of the key factors in estimating GFR is the serum creatinine level, which should be obtained through a blood test. 
	In some cases, blood tests and serum creatinine measurements may not be readily available.
	On the other hand, age is another important variable in GFR estimation and is easier to obtain. It is known that GFR tends to decrease with age \citep{o2007age}. In this study, age is used as the concomitant variable for GFR. In our dataset, the correlations between age and the negative GFR for non-CKD and CKD populations are $\rho_X=0.633$ and $\rho_Y=0.510$, respectively. 
	
	%However, we need to have information of GFR estimation requires serum creatinine level, blood, which might not be convenient. However, GFR tends to decrease as people get older. Therefore, we use age as a concomitant variable. The correlation between age and the negative GFR is ($\rho_X=0.633$ and $\rho_Y=0.510$). 
	
	%One way to measure is albumin (sort of protein) in urine, but it requires 24-hour time window. So the ratio of albumin and creatine in urine is used instead. 
	%GFR needs several factors. 
	
\begin{figure}[!ht]
	    \centering
	    \includegraphics[width=0.5\textwidth]{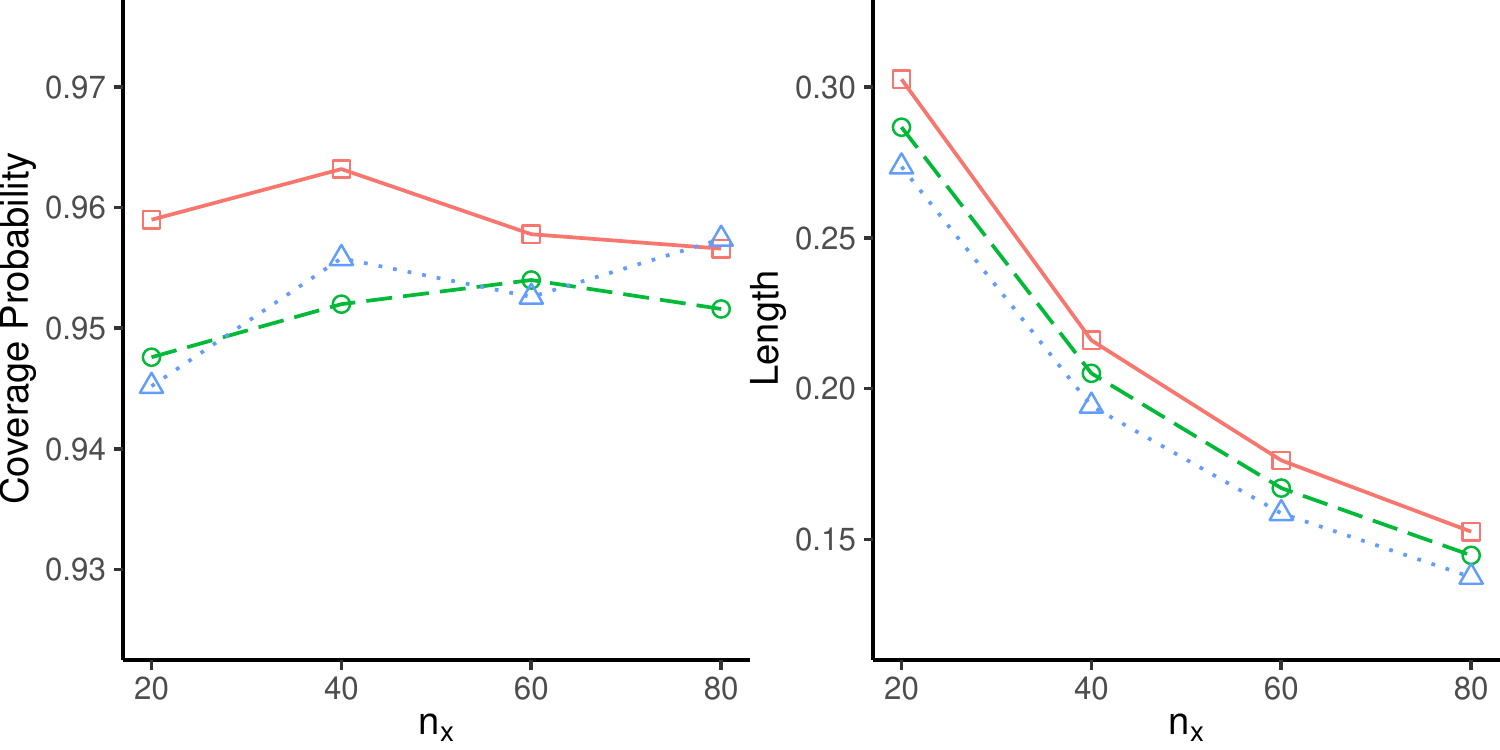}
	    \caption{Coverage probabilities and average lengths for 95\% confidence intervals of the AUC for the CKD data when age is used as the concomitant variable. Three lines are shown: SRS-EL as solid lines with square markers ($\square$), BRSS-EL with $m=2$ as long-dashed lines with circle markers ($\circ$), and BRSS-EL with $m=4$ as dotted lines with triangle markers ($\triangle$).}
	    \label{fig:ckd}
	\end{figure}

	We sample under the same setting of Section~\ref{subsec:5.1}: randomly generate samples 5,000 times from two sub-populations with sample sizes $n_x=n_y \in \left\{20,40,60,80 \right\}$, $m=n\in\left\{2,4\right\}$, and $k=n_x$/$m$ and $l=n_y$/$n$. Figure~\ref{fig:ckd} presents coverage probabilities and average lengths for 95\% confidence intervals of the AUC for SRS-EL, BRSS-EL with $m=2$, and BRSS-EL with $m=4$. 
	
	Using age as the concomitant variable in BRSS-EL leads to more efficient inference on the AUC than the other methods.
	BRSS-EL and SRS-EL achieve similar coverage probabilities close to $0.95$, while BRSS-EL has coverage probabilities closer to the nominal level. Also, the average lengths of the confidence intervals of BRSS-EL are shorter than SRS-EL regardless of the sample size.
	%BRSS-EL and SRS-EL achieve similar coverage probabilities close to 0.95. However, the average lengths of the confidence intervals of BRSS-EL are shorter than SRS-EL regardless of the sample size. This implies that using age as the concomitant variable in BRSS-EL enables more accurate inference on the AUC estimated by GFR.
	The BRSS-KER results are presented in Figure~\ref{fig:ckd.full} in Appendix~\ref{subapp:tbfig}. Although BRSS-KER has the shortest average lengths of the confidence intervals, its coverage probabilities are lower than the others. 
	
	\section{Discussion}
	\label{sec:conclusion}
	We propose the EL method that finds the confidence interval of the AUC using data obtained from balanced and unbalanced RSS. We show that the EL-based confidence intervals can be found by the scaled chi-square distributions. The proposed method performs the best in the sense that its confidence interval achieves the coverage probability close to the nominal level with the shorter length. Simulation studies show that the proposed method performs well regardless of distributions of diseased and non-diseased subjects and AUC values.
	We apply the proposed approach to diabetes and CKD data and show it outperforms the other methods.

	Our study suggests that the well-designed RSS may lead to more efficient inference on the AUC. 
	First, the proposed method shows improvement over SRS-EL regardless of the quality of ranking. Therefore, implementing RSS and estimating the confidence interval of the AUC using EL is a better strategy than SRS-EL.
	Second, one can obtain improved results by carefully assigning the number of ranked sets in URSS. 
	For example, simulation studies show that when the AUC value is close to one, taking a larger number of low-rank samples from the diseased population in URSS leads to improved inference.
	This could be useful in practice when some prior information of the AUC is available. 
	
	The proposed AUC estimator treats the X and Y samples in an asymmetric way; it uses the distribution function of non-diseased subjects $F$ as a reference and constructs the confidence interval for $P(X \leq Y)$. However, the AUC can also be estimated using the distribution of diseased subjects $G$ as the reference and then estimating the confidence interval for $1-P(Y<X)$. 
	Let the placement value of $X$ be $W_{ij} = G(X_{[i]j})$, a proportion of diseased subjects whose measurements are smaller than $X_{[i]j}$. The AUC can be obtained by $E\left[\sum\limits_{i=1}^m\sum\limits_{j=1}^{k} \frac{1}{mk}\left(1 - W_{ij} \right)\right]=E[1-G(X) ]=\delta$. The profile EL for the AUC of BRSS can be constructed as
	\begin{equation*}
	    L_{BRSS}^*(\delta_0) = \sup \left\{ \prod\limits_{i=1}^{m}\prod\limits_{j=1}^{k} q_{ij}: \sum\limits_{i=1}^{m}\sum\limits_{j=1}^{k} q_{ij}=1, \sum\limits_{i=1}^m\sum\limits_{j=1}^{k} q_{ij}  \left(1-W_{ij}-\delta_0\right) =0 \right\},
	\end{equation*}
	where $\sum_{i=1}^{m}\sum_{j=1}^{k} q_{ij}=1$ and $q_{ij}\geq0$ for all $i=1,\ldots,m$ and $j=1,\ldots,k$.
	Similar to Section~\ref{sec:method}, $W_{ij}$ can be estimated by the empirical distribution $\widehat{W}_{ij}=\widehat{G}\left(X_{[i]j}\right)=\frac{1}{nl}\sum\limits_{r=1}^n\sum\limits_{s=1}^l\phi\left(Y_{[r]s},X_{[i]j}\right)$ and the empirical log-likelihood ratio becomes 
	\begin{equation*}
	    l_{BRSS}^*(\delta_0) = 2\sum\limits_{i=1}^m \sum\limits_{j=1}^k \log \left( 1+ \lambda \left(1-\widehat{W}_{ij}-\delta_0\right) \right).
	\end{equation*}
	The asymptotic distribution of $l_{BRSS}^*(\delta_0)$ also follows the scaled chi-square distribution
	\begin{equation*}
	    r^*_{BRSS}(\delta_0)l_{BRSS}^*(\delta_0) \rightarrow \chi^2_1, \text{ where}
	\end{equation*}
	\begin{equation*}
	    r^*_{BRSS}(\delta_0)=\frac{nl}{mk+nl} \frac{\sum\limits_{i=1}^m \sum\limits_{j=1}^{k} \frac{1}{mk} \left(1- \widehat{W}_{ij}-\delta_0 \right)^2}{S_{BRSS}^2}.
	\end{equation*}
	A similar extension can also be applied to URSS. In practice, it would be natural to set $F$ as the reference because the number of the non-diseased sample $n_x$ is expected to be larger than the number of the diseased sample $n_y$. Table~\ref{tb:cis} compares the coverage probabilities and average lengths of the AUC for two approaches when $X$ and $Y$ are sampled using BRSS from the normal distributions presented in Section~\ref{subsec:4-1} with $n_x\in\{40,100,200\}$, $n_y=40$, $m=n=2$, $\rho=1$, and $\delta=0.8$. 
	%We compare confidence intervals of two approaches using simulations. The non-diseased and diseased subjects $X$ and $Y$ are sampled from the normal distributions in Section~\ref{subsec:4-1} with $n_x=n_y=40$, $m=n=2$, $\rho=1$, and $\delta\in{0.6,0.8,0.9}$. The BRSS samples are generated 1,000 times for $X$ and $Y$, respectively.
	As $n_x$ gets larger, the confidence interval of using $F$ as the reference becomes narrower while maintaining the coverage probability close to the nominal level. On the other hand, the length of the confidence interval using $G$ as the reference has little change, but its standard deviation gets larger as $n_x$ increases.

	%In the future, we plan to develop JEL methods for AUC with RSS. \cite{Jing2009} show that the confidence interval of AUC computed by Jackknife EL (JEL) can achieve higher coverage probability than the scaled EL of \cite{Qin2006} at the cost of longer intervals. 
	%Also, \cite{Zhang2016} apply JEL to RSS for estimating population means. 
	
	An interesting future work is using the nonparametric maximum likelihood estimator (NPMLE) $F^*$ of Kvam and Samaniego\cite{Kvam1994} instead of the empirical distribution $\widehat{F}$ of Stokes and Sager\cite{Stokes1988}. The NPMLE assigns different probabilities to observations and is known to perform better than the empirical distribution. Thus, implementing the NPMLE could further improve the efficiency of inference on the AUC. 
	Another interesting topic is to determine the optimal number of set sizes for the AUC estimation. A few studies have investigated the optimal set size for one-population problems by considering the quality of judgment ranking, the impact of imperfect rankings, and cost of sampling\cite{nahhas2002ranked,buchanan2005cost}. However, for two-population problems such as the AUC estimation, the cost of recruiting units from one sub-population might be much more than that from the other. Thus, it might be worth a formal investigation in the future.

	\section*{Conflict of interest}
The authors declare no potential conflict of interests.
	
	\section*{Data Availability Statement}
	All data used in simulation studies are generated randomly and the NHANES data used in the case studies are imported from \url{https://www.cdc.gov/nchs/nhanes/}. The R code used to generate, import, and analyze the data used in this paper is publicly available at the URL:\url{https://github.com/chulmoon/EL-RSS-AUC}.

	%\section*{Declaration of Interests}
	%Declarations of interest: none
	
	%\section*{Financial Support}
	%The authors received no financial support for this article. 
	
	%\section*{CRediT Author Statement}
	%\textbf{Chul Moon:} Conceptualization, Methodology, Software, Formal analysis, Writing - Original Draft. \textbf{Xinlei Wang:} Conceptualization, Writing - Review \& Editing. \textbf{Johan Lim:} Writing - Review \& Editing.

	\appendix
	\label{sec:appendix}
	\section{Proof of Theorems}
	\label{subapp:proof}
	\begin{lemma}
	Under the assumptions of Theorem~\ref{theorem1}, 
		\begin{eqnarray*}
			\text{(i)} &&  \sum\limits_{r=1}^n\sum\limits_{s=1}^l \frac{1}{nl} \left( 1- \widehat{U}_{rs}-\delta_0 \right)^2 \xrightarrow{p} 
			\frac{1}{n}\left( \sum\limits_{r=1}^n\left(\sigma_{0[r]}^Y\right)^2 
			+ \sum\limits_{r=1}^n\left(\delta_{0[r]}^Y-\delta_0 \right)^2
			\right), \\
			&&\text{where } \sigma_{[r]}^Y=E\left[ \left( F(Y_{[r]})-\delta_{0[r]}^Y\right)^2 \right], \\
			&&\text{and }  \delta_{0[r]}^Y \text{ is the true AUC when the $r$th order statistic from a SRS of size $n$ of diseased subjects is used}.
			%\sigma_0^2 \text{, where } 
			%\sigma^2_0= E\left[F^2(Y)\right]-\delta_{0}^2
			\\
			%\frac{1}{n}\left( F^2(Y_{[r]}) \right) - \delta_0^2\\
			\text{(ii)} && \left( \frac{mknl}{mk+nl} \right)^{1/2} \frac{\sum\limits_{r=1}^n\sum\limits_{s=1}^l \frac{1}{nl} \left(1-\widehat{U}_{rs} - \delta_0\right)}{S_{BRSS}} \xrightarrow{D} N(0,1).
		\end{eqnarray*}
		 \label{lemma1}
	\end{lemma}
	
	\begin{proof}
		(i) Using the uniform consistency of the empirical distribution $\widehat{F}$ we have,
		\begin{eqnarray*}
		\sum\limits_{r=1}^n\sum\limits_{s=1}^l \frac{1}{nl} \left( 1- \widehat{U}_{rs}-\delta_0 \right)^2 &=& \sum\limits_{r=1}^n\sum\limits_{s=1}^l \frac{1}{nl} \left( \widehat{F}(Y_{[r]s})-\delta_0 \right)^2 \\
		&=& \sum\limits_{r=1}^n\sum\limits_{s=1}^l \frac{1}{nl}  \left( \widehat{F}(Y_{[r]s})-\delta_{0[r]}^Y+\delta_{0[r]}^Y-\delta_0 \right)^2 \\
		&=& \sum\limits_{r=1}^n\sum\limits_{s=1}^l \frac{1}{nl}  \left( \widehat{F}(Y_{[r]s})-\delta_{0[r]}^Y\right)^2+\sum\limits_{r=1}^n\sum\limits_{s=1}^l \frac{1}{nl}\left(\delta_{0[r]}^Y-\delta_0 \right)^2 \\
		&\xrightarrow{p}& \frac{1}{n}\left( \sum\limits_{r=1}^n\left(\sigma_{[r]}^Y\right)^2 
			+ \sum\limits_{r=1}^n\left(\delta_{0[r]}^Y-\delta_0 \right)^2
			\right).
			%\text{ as }l\rightarrow\infty.
		%&\xrightarrow{p}&  \frac{1}{n}\left( F^2(Y_{[r]}) \right) - \delta_0^2 \text{ as }l\rightarrow\infty.
		\end{eqnarray*}
		%Lemma \ref{lemma1}(i) is proved because . 
		
		(ii)  %Lemma \ref{lemma1}(ii) is an extension of Theorem 3.1 of \cite {Sen1967} to RSS. 
		For RSS, the distribution functions of diseased and non-diseased subjects can be written as $G(x)=\frac{1}{n}\sum\limits_{r=1}^n G_{[r]}(x)$ and $F(x)=\frac{1}{m}\sum\limits_{i=1}^m F_{[i]}(x)$, respectively, regardless of the quality of the judgement ranking \citep{Presnell1999}. 
		%only $X_{[i]j}$'s are the same, only $Y_{[r]}s$'s are the same, both $X_{[i]j}$'s and $Y_{[r]}s$'s are the same, none of $X_{[i]j}$'s and $Y_{[r]}s$'s are the same.
		By simple calculations,
		\begin{eqnarray*}
		\text{Var}(\widehat{\delta}_{BRSS}) &=& \frac{1}{mknl}\left\{  
		nl\left(\beta  - \frac{1}{m}\sum\limits_{i=1}^m \left(\delta_{[i]}^X \right)^2  \right) 
		+ mk\left(\alpha  - \frac{1}{n}\sum\limits_{r=1}^n \left(\delta_{[r]}^Y \right)^2  \right) \right. \nonumber  \\
		&& \left. + \left( \delta_0 - \overline{\delta}_0^2 - \left(\overline{\beta} - \overline{\delta}_0^2\right) - \left(\overline{\alpha} - \overline{\delta}_0^2\right) \right)  \right\},
		\end{eqnarray*}
		where $\beta=\int_{0}^1\left(1-G(x)\right)^2dF(x)$, $\overline{\beta}=\int_{0}^1 \frac{1}{n}\sum\limits_{r=1}^n \left(1-G_{[r]}(x)\right)^2dF(x)$,
		$\alpha=\int_{0}^1 F^2(x)dG(x)$, 
		$\overline{\alpha}=\int_{0}^1 \frac{1}{m}\sum\limits_{i=1}^m F_{[i]}^2(x)dG(x)$, $\delta_{[r]}^Y=P(X \leq Y_{[r]} )$, $\delta_{[i]}^X=P(X_{[i]} \leq Y)$, and $\overline{\delta}_0^2 = \frac{1}{mn}\sum\limits_{i=1}^m\sum\limits_{r=1}^n \left( P(X_{[i]}<Y_{[r]}) \right)^2 $.
		
		%Let 
		%\begin{eqnarray*}
		%	V^{10}(X_{[i]j}) &=& \frac{1}{nl}\sum\limits_{r=1}^{n}\sum\limits_{s=1}^l \phi(X_{[i]j},Y_{[r]s}) \\
		%	V^{01}(Y_{[r]s}) &=& \frac{1}{mk}\sum\limits_{i=1}^{m}\sum\limits_{j=1}^k \phi(X_{[i]j},Y_{[r]s}).
		%\end{eqnarray*}
		
		%Let further
		%\begin{eqnarray}
		%\left(S^{10}_{[i]}\right)^2 &=& \frac{1}{k-1} \sum\limits_{j=1}^k \left( V^{10}(X_{[i]j})- \overline{V}_{[i]}^{10} \right)^2 \label{eq:5} \\
		%\left(S^{01}_{[r]}\right)^2 &=& \frac{1}{l-1} \sum\limits_{s=1}^l \left( V^{01}(Y_{[r]s})-\overline{V}_{[r]}^{01} \right)^2. \label{eq:6}
		%\end{eqnarray}
		
		%where $\overline{V}_{[i]}^{10}=\frac{1}{k}\sum\limits_{j=1}^k V^{10}(X_{[i]j})$ and $\overline{V}_{[r]}^{01}=\frac{1}{l}\sum\limits_{s=1}^l V^{01}(Y_{[r]s})$,
		
		%\begin{eqnarray}
		%\left(S^{10}\right)^2 = \frac{1}{m}\sum\limits_{i=1}^m \left(S^{10}_{[i]}\right)^2,\\
		%\left(S^{01}\right)^2 = \frac{1}{n}\sum\limits_{r=1}^n \left(S^{01}_{[r]}\right)^2.
		%\end{eqnarray}

		Then, 
		\begin{eqnarray*}
			nlE\left[ \left(S_{BRSS}^{10}\right)^2 \right] &=&   \sum\limits_{i=1}^{m}\sum\limits_{j=1}^k \frac{nl}{m(k-1)} E\left[ V_{BRSS}^{10}(X_{[i]j})V_{BRSS}^{10}(X_{[i]j}) \right]  - \sum\limits_{i=1}^{m} \frac{knl}{m(k-1)} E\left[ \overline{V}_{BRSS[i]}^{10}\overline{V}_{BRSS[i]}^{10} \right] \\
			&=& nl\left(\beta - \frac{1}{m}\sum\limits_{i=1}^m \left(\delta_{[i]}^X \right)^2  \right) - \overline{\beta} - \overline{\alpha} + \delta_0 - \overline{\delta}_0^2, \\
			mkE\left[ \left(S_{BRSS}^{01}\right)^2 \right] &=&   \sum\limits_{r=1}^{n}\sum\limits_{s=1}^l \frac{mk}{n(l-1)} E\left[ V_{BRSS}^{01}(Y_{[r]s}) V_{BRSS}^{01}(Y_{[r]s}) \right] -    \sum\limits_{r=1}^{n} \frac{mkl}{n(l-1)} E \left[ \overline{V}_{BRSS[r]}^{01}\overline{V}_{BRSS[r]}^{01} \right]  \\
			&=& mk\left(\alpha -\frac{1}{n}\sum\limits_{r=1}^n \left(\delta_{[r]}^Y \right)^2  \right) - \overline{\alpha} - \overline{\beta} + \delta_0 - \overline{\delta}_0^2. 
		\end{eqnarray*}
		
		Therefore, 
		\begin{eqnarray*}
			E\left[S_{BRSS}^2\right]&=&E\left[\frac{nl\left(S_{BRSS}^{10}\right)^2+mk\left(S_{BRSS}^{01}\right)^2}{mk+nl}\right] \\
			&=& \text{Var}\left(\left(\frac{mknl}{mk+nl}\right)^{\frac{1}{2}}(\widehat{\delta}_{BRSS}-\delta_0)\right)+ \frac{\delta_0 - \overline{\delta}_0^2 - \left(\overline{\beta}- \overline{\delta}_0^2\right)   - \left(\overline{\alpha}-\overline{\delta}_0^2\right) }{mk+nl}.
		\end{eqnarray*}
		
		Now, we will show that the bias of $S_{BRSS}^2$ is less than $\frac{\delta_0 - \overline{\delta}_0^2 }{mk+nl}$.
		Let the counter function be
		\begin{eqnarray*}
			\overline{\phi}(X_{[i]j},Y_{[r]s}) = \phi(X_{[i]j},Y_{[r]s})-\overline{\delta}_{ir}-
			\left(\phi_{10}(X_{[i]j})-\overline{\delta}_{i r}\right)-\left(\phi_{01}(Y_{[r]s})-\overline{\delta}_{i r}\right),
		\end{eqnarray*}
		where $\phi_{10}(x_{[i]j})=\sum\limits_{r=1}^n\sum\limits_{s=1}^l \frac{1}{nl} E\left[ \phi(x_{[i]j},Y_{[r]s}) \right]$, $\phi_{01}(y_{[r]s})=\sum\limits_{i=1}^m\sum\limits_{j=1}^k \frac{1}{mk} E\left[ \phi(X_{[i]j},y_{[r]s})\right]$ and $\overline{\delta}_{ir} = P(X_{[i]}<Y_{[r]})$. 
		
		Then,
		\begin{eqnarray*}
			&&\frac{1}{mknl}\sum\limits_{i=1}^m\sum\limits_{j=1}^k\sum\limits_{r=1}^n\sum\limits_{s=1}^l E \left[ \overline{\phi}^2(X_{[i]j},Y_{[r]s}) \right]\\ 
			&=& \frac{1}{mknl}\sum\limits_{i=1}^m\sum\limits_{j=1}^k\sum\limits_{r=1}^n\sum\limits_{s=1}^l E \left[ \left( \phi(X_{[i]j},Y_{[r]s})-\overline{\delta}_{ir}-
			\left(\phi_{10}(X_{[i]j})-\overline{\delta}_{i r}\right)
			-\left(\phi_{01}(Y_{[r]s})-\overline{\delta}_{i r}\right) \right)^2 \right] \\
			&=& \delta_0-\overline{\delta}_0^2 - \left(\overline{\alpha}-\overline{\delta}_0^2\right) - \left(\overline{\beta}-\overline{\delta}_0^2\right) \geq 0.
		\end{eqnarray*}
		Because
		\begin{eqnarray*}
			&&\frac{1}{mknl}\sum\limits_{i=1}^m\sum\limits_{j=1}^k\sum\limits_{r=1}^n\sum\limits_{s=1}^l 
			E\left[\left(\phi_{10}(X_{[i]j})-\overline{\delta}_{ir}\right)^2\right] = \overline{\alpha}-\overline{\delta}_0^2 \geq 0, \\
			&&\frac{1}{mknl}\sum\limits_{i=1}^m\sum\limits_{j=1}^k\sum\limits_{r=1}^n\sum\limits_{s=1}^l 
			E\left[\left(\phi_{01}(Y_{[r]s})-\overline{\delta}_{i r}\right)^2\right] = \overline{\beta}-\overline{\delta}_0^2 \geq 0,
		\end{eqnarray*}
		it implies 
		\begin{equation}
		0 \leq \left(\overline{\alpha}-\overline{\delta}_0^2\right)+\left(\overline{\beta}-\overline{\delta}_0^2\right)  \leq \delta_0-\overline{\delta}_0^2.
		\label{eq:7}
		\end{equation}
		%Also, because $\frac{1}{n}\sum\limits_{r=1}^n (1-G_{[r]}(x)) = 1-G(x)$ and $\frac{1}{m}\sum\limits_{i=1}^m F_{[i]}(x) = F(x)$, by applying Cauchy-Schwarz inequality, we have $\overline{\beta} \geq \beta$ and $\overline{\alpha} \geq \alpha$. 
		%\begin{eqnarray*}
		%    \frac{1}{n}\sum\limits_{r=1}^n\left(\delta_{[r]}^Y\right)^2 \geq \delta^2_0  \\ 
		%    \frac{1}{m}\sum\limits_{i=1}^m \left(\delta_{[i]}^X\right)^2 \geq \delta_0^2,
		%\end{eqnarray*}
		
		%Therefore,
		%\begin{equation}
		%    0 \leq (\overline{\alpha} - \alpha) +  (\overline{\beta}-\beta)
		%      \leq \delta_0(1-\delta_0)
		%      \label{eq:8}
		%\end{equation}
		By (\ref{eq:7}), we get the upper bound of the bias of $S^2$.
		
		Let 
		\begin{eqnarray*}
			&&T(X_{[1]1}, \ldots, X_{[m]k},Y_{[1]1},\ldots,Y_{[n]l}) \\
			&&= \frac{1}{mk}\sum\limits_{i=1}^m\sum\limits_{j=1}^k \left( \phi_{10}(X_{[i]j}) -\delta_{[i]}^X \right) + \frac{1}{nl}\sum\limits_{r=1}^n\sum\limits_{s=1}^l \left( \phi_{01}(Y_{[r]s})-\delta_{[r]}^Y \right).
		\end{eqnarray*}
		
		Then,
		\begin{eqnarray*}
			&&\frac{mknl}{mk+nl}E\left[ \left\{ T(X_{[1]1}, \ldots, X_{[m]k},Y_{[1]1},\ldots,Y_{[n]l}) - (\widehat{\delta}_{BRSS}-\delta_0) \right\}^2  \right]\\ 
			&=& \frac{\delta_0-\overline{\delta}_0^2 - \left(\overline{\alpha}-\overline{\delta}_0^2\right) - \left(\overline{\beta}-\overline{\delta}_0^2\right)}{mk+nl}\\
			%&&+\left(\frac{1}{n}\sum\limits_{r=1}^n\left(\delta_{[r]}^Y\right)^2 - \delta^2_0 \right)\frac{mk}{mk+nl}  + 
			%\left( \frac{1}{m}\sum\limits_{i=1}^m \left(\delta_{[i]}^X\right)^2 - \delta^2_0 \right) \frac{nl}{mk+nl}\\
			&\leq& \frac{\delta_0-\overline{\delta}_0^2}{mk+nl} \\
			%+\left(\frac{1}{n}\sum\limits_{r=1}^n\left(\delta_{[r]}^Y\right)^2 - \delta^2_0 \right)\frac{mk}{mk+nl}  + 
			%\left( \frac{1}{m}\sum\limits_{i=1}^m \left(\delta_{[i]}^X\right)^2 - \delta^2_0 \right) \frac{nl}{mk+nl} \\
			&=& O( \max(k,l)^{-1} ).
		\end{eqnarray*}
	
		By Chebyshev's inequality, 
		\begin{equation}
		\frac{mknl}{mk+nl}\left(\widehat{\delta}_{BRSS}-\delta_0\right) \xrightarrow{p} \frac{mknl}{mk+nl}T\left(X_{[1]1}, \ldots, X_{[m]k},Y_{[1]1},\ldots,Y_{[n]l}\right).
		\label{eq:convp}
		\end{equation}
		As $k\rightarrow\infty$ and $l\rightarrow\infty$, by the central limit theorem, 
		\begin{eqnarray}
		\frac{1}{\sqrt{mk}}\sum\limits_{i=1}^m\sum\limits_{j=1}^k \phi_{10}(X_{[i]j}) &\xrightarrow{D}& N\left(0,\overline{\beta}-\overline{\delta}_0^2\right) \label{eq:convd1} \\
		\frac{1}{\sqrt{nl}}\sum\limits_{r=1}^n\sum\limits_{s=1}^l \phi_{01}(Y_{[r]s}) &\xrightarrow{D}& N\left(0,\overline{\alpha}-\overline{\delta}_0^2\right). \label{eq:convd2}
		\end{eqnarray}
		Thus from (\ref{eq:convp}), (\ref{eq:convd1}), (\ref{eq:convd2}), and $S_{BRSS}^2  \xrightarrow{p} \text{Var}\left(\frac{mknl}{mk+nl}\left(\widehat{\delta}_{BRSS}-\delta_0\right)\right)$,  Lemma~\ref{lemma1}(ii) is proved. 
		%The formulas for $S^2_{10}$ and $S^2_{01}$ in Lemma~\ref{lemma1} are algebraic expressions of (\ref{eq:5}) and (\ref{eq:6}). 
	\end{proof}

	\begin{proof}[Proof of Theorem~\ref{theorem1}]
	Using Lemma~\ref{lemma1} and the similar arguments of Owen\cite{Owen1990}, we can show that $|\lambda|=O_p(n_y^{-1/2})$ and
		\begin{eqnarray}
		\lambda = \frac{\sum\limits_{r=1}^n \sum\limits_{s=1}^l \left(1-\widehat{U}_{rs}-\delta_0\right)}{\sum\limits_{r=1}^n \sum\limits_{s=1}^l \left( 1-\widehat{U}_{rs}-\delta_0 \right)^2} + O_p(n_y^{-1/2}).
		\label{eq:9}
		\end{eqnarray}
		
		By applying Taylor expansion to equation~(\ref{eq:4}) in Section~\ref{subsec:2.2}, 
		\begin{eqnarray*}
			l(\delta_0) = 2\sum\limits_{r=1}^n \sum\limits_{s=1}^l \left( \lambda \left(1-\widehat{U}_{rs}-\delta_0\right) - \frac{1}{2}\left( \lambda \left(1-\widehat{U}_{rs}-\delta_0\right) \right)^2 \right) + R_{n_y},
		\end{eqnarray*}
		where $R_{n_y} \leq C \sum\limits_{r=1}^n \sum\limits_{s=1}^l \left|\lambda \left(1-\widehat{U}_{rs}-\delta_0\right)\right|^3 $ for some constant $C$. Because $F\in[0,1]$ and so the third moment of $F(Y_{[r]s})$ is finite, and $|\lambda|=O_p(n_y^{-1/2})$,
		\begin{eqnarray*}
			\left|R_{n_y}\right| \leq 
			C \sum\limits_{r=1}^n \sum\limits_{s=1}^l \left|\lambda \left(1-\widehat{U}_{rs}-\delta_0\right)\right|^3  =  
			C \sum\limits_{r=1}^n \sum\limits_{s=1}^l \left|\lambda \left(\widehat{F}(Y_{[r]s})-\delta_0\right)\right|^3
			\leq
			C \left|\lambda\right|^3n_y  = O_p(n_y^{-1/2}).
		\end{eqnarray*}
		%\begin{eqnarray*}
		%	|R_{n_y}| = O_p(n_y^{-1/2}).
		%\end{eqnarray*}
		Also, equation~(\ref{eq:3}) in Section~\ref{subsec:2.2} can be expressed as,
		\begin{eqnarray}
		0 &=&  \sum\limits_{r=1}^n \sum\limits_{s=1}^l \frac{  \left(1-\widehat{U}_{rs}-\delta_0\right) }{1+\lambda \left(1-\widehat{U}_{rs}-\delta_0\right)} \nonumber \\
		&=&  \sum\limits_{r=1}^n \sum\limits_{s=1}^l \lambda\left(1-\widehat{U}_{rs}-\delta_0\right) \left(1-\lambda\left(1-\widehat{U}_{rs}-\delta_0\right) + \frac{ \left(\lambda\left(1-\widehat{U}_{rs}-\delta_0\right)\right)^2}{1+ \lambda\left(1-\widehat{U}_{rs}-\delta_0\right)} \right) \nonumber \\
		&=&  \sum\limits_{r=1}^n \sum\limits_{s=1}^l \lambda\left(1-\widehat{U}_{rs}-\delta_0\right) - \sum\limits_{r=1}^n \sum\limits_{s=1}^l \left(\lambda\left(1-\widehat{U}_{rs}-\delta_0\right)\right)^2 + \sum\limits_{r=1}^n \sum\limits_{s=1}^l  \frac{ \left(\lambda\left(1-\widehat{U}_{rs}-\delta_0\right)\right)^3}{1+ \lambda\left(1-\widehat{U}_{rs}-\delta_0\right)} \nonumber \\
		&=&  \sum\limits_{r=1}^n \sum\limits_{s=1}^l \lambda\left(1-\widehat{U}_{rs}-\delta_0\right) - \sum\limits_{r=1}^n \sum\limits_{s=1}^l \left(\lambda\left(1-\widehat{U}_{rs}-\delta_0\right)\right)^2 + O_p(n_y^{-1/2}), \label{eq:10}
		\end{eqnarray}
		%using similar arguments of Owen\cite{Owen1990}. 
		
		Therefore, 
		\begin{eqnarray*}
			r(\delta_0)l(\delta_0) &=&  2r(\delta_0)\sum\limits_{r=1}^n \sum\limits_{s=1}^l \left( \lambda\left(1-\widehat{U}_{rs}-\delta_0\right) -\frac{1}{2}(\lambda\left(1-\widehat{U}_{rs}-\delta_0\right))^2 \right) + R_{n_y} \\
			&=& r(\delta_0)\left( 2 \sum\limits_{r=1}^n \sum\limits_{s=1}^l  \lambda\left(1-\widehat{U}_{rs}-\delta_0\right) - \sum\limits_{r=1}^n \sum\limits_{s=1}^l \left( \lambda\left(1-\widehat{U}_{rs}-\delta_0\right) \right)^2 \right) + O_p(n_y^{-1/2}) \\
			&=& r(\delta_0)\left( \sum\limits_{r=1}^n \sum\limits_{s=1}^l \left( \lambda\left(1-\widehat{U}_{rs}-\delta_0\right) \right)^2 \right) + O_p(n_y^{-1/2}) \text{, using (\ref{eq:10})}\\
			&=& r(\delta_0) \frac{ \left(\sum\limits_{r=1}^n \sum\limits_{s=1}^l \left(1-\widehat{U}_{rs}-\delta_0\right) \right)^2}{\sum\limits_{r=1}^n \sum\limits_{s=1}^l \left((1-\widehat{U}_{rs}-\delta_0 \right)^2}   + O_p(n_y^{-1/2}) \text{, by plugging (\ref{eq:9})} \\
			&=& \frac{mknl}{mk+nl}\frac{ \left(\sum\limits_{r=1}^n \sum\limits_{s=1}^l \frac{1}{nl} \left(1-\widehat{U}_{rs}-\delta_0\right) \right)^2}{S^2}  + O_p(n_y^{-1/2}) \\
			&=& \left[ \left(\frac{mknl}{mk+nl}\right)^{1/2} \frac{  \sum\limits_{r=1}^n \sum\limits_{s=1}^l \frac{1}{nl} \left(1-\widehat{U}_{rs}-\delta_0\right) }{S} \right]^2  + O_p(n_y^{-1/2}) \\
			&\xrightarrow{D}& \chi^2_1.
		\end{eqnarray*}
	\end{proof}
	
	\begin{lemma}
	Under the assumptions of Theorem~\ref{theorem2},
		\begin{eqnarray*}
			\text{(i)} &&  \sum\limits_{r=1}^n\sum\limits_{s=1}^{l_r} \frac{1}{nl_r} \left( 1- \widehat{U}_{rs}-\delta_0 \right)^2 \xrightarrow{p} 
			\frac{1}{n}\left( \sum\limits_{r=1}^n\left(\sigma_{0[r]}^Y\right)^2 
			+ \sum\limits_{r=1}^n\left(\delta_{0[r]}^Y-\delta_0 \right)^2
			\right), \\
			&&\text{where } \sigma_{[r]}^Y=E\left[ \left( F(Y_{[r]})-\delta_{0[r]}^Y\right)^2 \right], \\
			&&\text{and }  \delta_{0[r]}^Y \text{ is the true AUC when the $r$th order statistic from a SRS of size $n$ of diseased subjects is used}.
			%\sigma_0^2 \text{, where } 
			%\sigma^2_0= E\left[F^2(Y)\right]-\delta_{0}^2
			\\
			%\frac{1}{n}\left( F^2(Y_{[r]}) \right) - \delta_0^2\\
			\text{(ii)} && \left( \frac{n_x n_y}{n_x+n_y} \right)^{1/2} \frac{\sum\limits_{r=1}^n\sum\limits_{s=1}^{l_r} \frac{1}{nl_r} \left(1-\widehat{U}_{rs} - \delta_0\right)}{S} \xrightarrow{D} N(0,1).
		\end{eqnarray*}
		 \label{lemma2}
	\end{lemma}
	
    \begin{proof}
    \noindent The proof is a simple modification of Lemma~\ref{lemma1}. By replacing $l$ with $l_r$, $k$ with $k_i$, $nl$ with $n_y$, $mk$ with $n_x$ and following the similar steps, Lemma~\ref{lemma2} can be proved.
    \end{proof}
    
    \begin{proof}[Proof of Theorem~\ref{theorem2}]
    \noindent Using Lemma~\ref{lemma2} and similar steps used in Theorem~\ref{theorem1}, Theorem~\ref{theorem2} can be proved. 
	\end{proof}

    \section{Tables and Figures}
    \label{subapp:tbfig}

\begin{center}
\begin{table*}[!ht]%
\caption{Comparison of the computation time. The total execution time for 5,000 replicates in each setting is reported. \label{table:time}}
\centering
\begin{tabular*}{500pt}{@{\extracolsep\fill}lccc@{\extracolsep\fill}}
\toprule
&\multicolumn{3}{@{}c@{}}{\textbf{Time (seconds)}} \\\cmidrule{2-4}
\textbf{Methods} & \textbf{$m=n=2$}  & \textbf{$m=n=4$}  & {\textbf{$m=n=5$}}  \\
\midrule
SRS-EL & 531 & 532  & 532     \\
BRSS-EL & 577  & 480  & 453    \\
BRSS-KER & 135  & 140  & 144  \\
\bottomrule
\end{tabular*}
\end{table*}
\end{center}

\begin{center}
    \begin{table}[!ht]
    \caption{Coverage probabilities and average lengths for 95\% confidence intervals of the AUC for two approaches: 1) using $F$ as the reference ($P(X\leq Y)$) and 2) using $G$ as the reference ($1-P(Y<X)$). The BRSS samples are generated 1,000 times for $X$ and $Y$, respectively.}\label{tb:cis}
\begin{tabular*}{500pt}{@{\extracolsep\fill}llll@{\extracolsep\fill}}
\toprule
%&\multicolumn{3}{@{}c@{}}{\textbf{Time (seconds)}}
                                    Sample size        & Reference distribution  &        Length (standard deviation)        & Coverage probability \\ \midrule
$n_x=40$, $n_y=40$ & Non-diseased, $F$  & 0.176 (0.018) & 0.942                \\ \cmidrule{2-4} 
                     & Diseased, $G$  & 0.187 (0.027) & 0.948                \\ \midrule
$n_x=100$, $n_y=40$ & Non-diseased, $F$  & 0.168 (0.017) & 0.950                \\ \cmidrule{2-4} 
                     & Diseased, $G$ & 0.187 (0.032) & 0.966                \\ \midrule
$n_x=200$, $n_y=40$ & Non-diseased, $F$   & 0.164 (0.018) & 0.945                \\ \cmidrule{2-4} 
                     & Diseased, $G$  & 0.186 (0.034) & 0.963                \\ \bottomrule
\end{tabular*}
\end{table}
\end{center}

	\begin{figure}[!ht]
        \centerline{\includegraphics[width=0.65\textwidth]{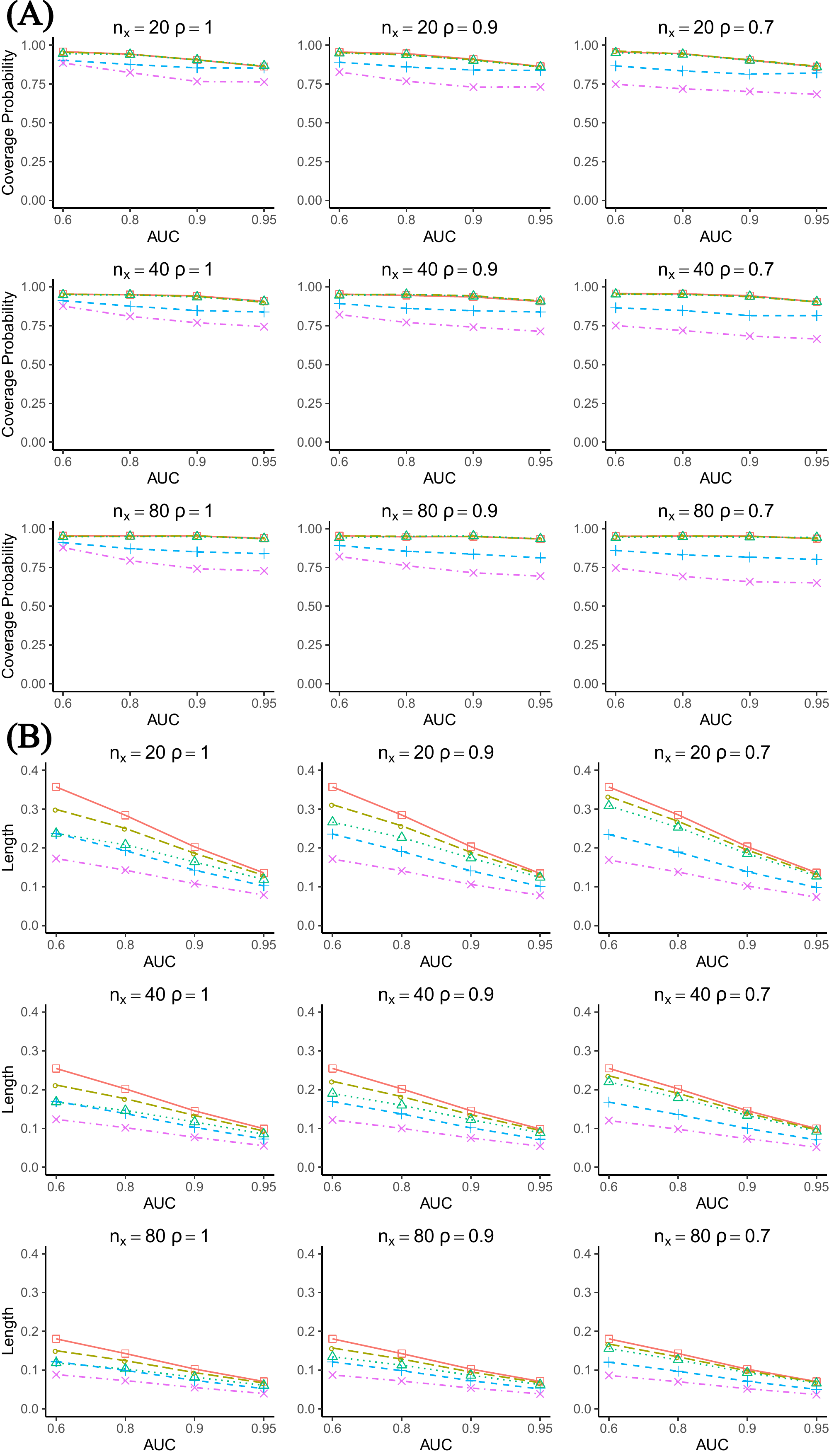}}
	    \caption{Estimated 95\% confidence intervals of the AUC for normal distributions: (A) coverage probabilities, (B) average lengths. Five lines are shown: SRS-EL as solid lines with square markers ($\square$), BRSS-EL with $m=2$ as long-dashed lines with circle markers ($\circ$), BRSS-EL with $m=4$ as dotted lines with triangle markers ($\triangle$), BRSS-KER with $m=2$ as dashed lines with plus markers ($+$), and BRSS-KER with $m=4$ as dash-dot lines with cross markers ($\times$).}
	    \label{fig:normal.cp.full}
	\end{figure}
	
	\begin{figure}[!ht]
	    \centerline{\includegraphics[width=0.65\textwidth]{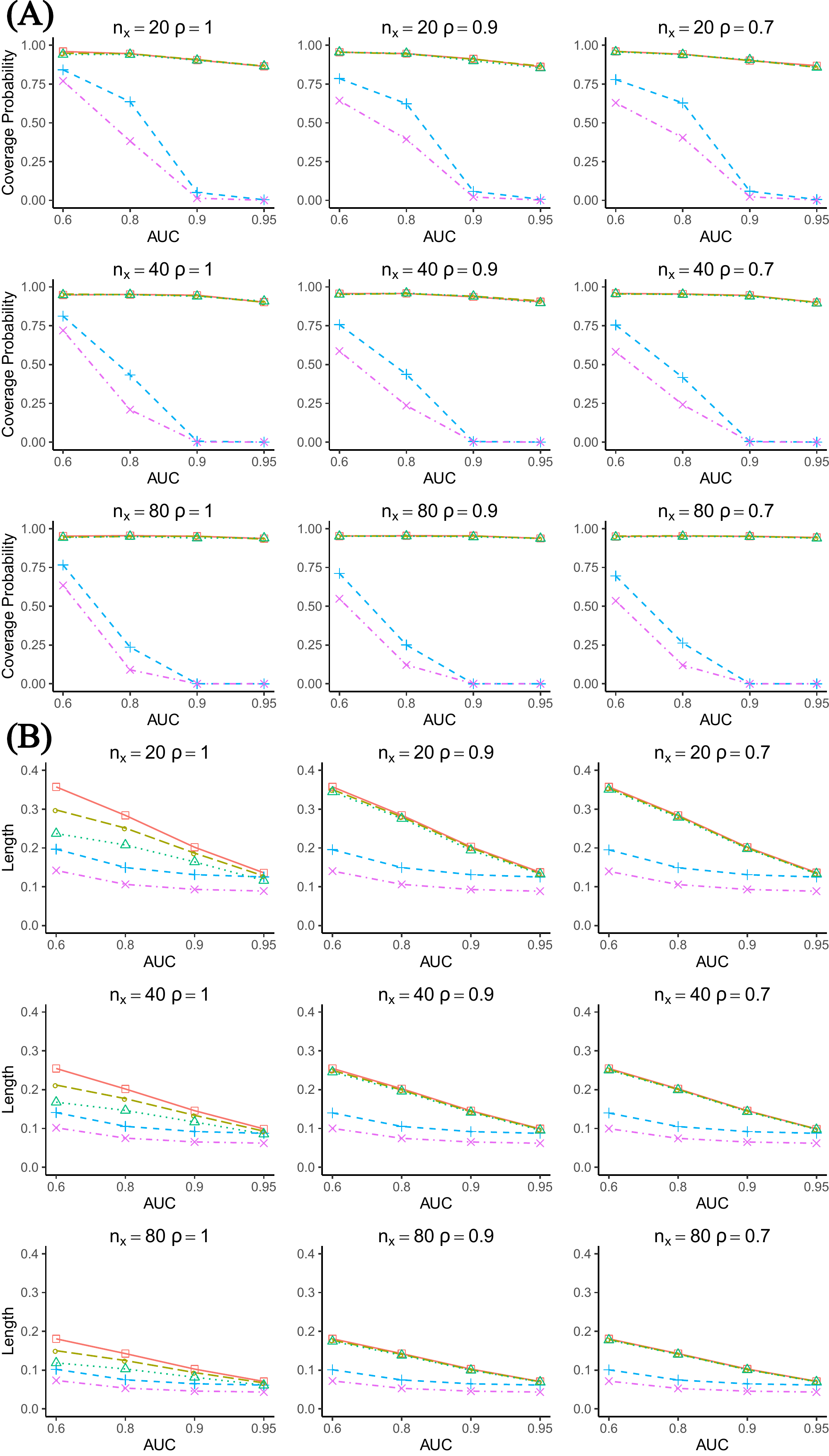}}
	    \caption{Estimated 95\% confidence intervals of the AUC for log-normal distributions: (A) coverage probabilities, (B) average lengths. Five lines are shown: SRS-EL as solid lines with square markers ($\square$), BRSS-EL with $m=2$ as long-dashed lines with circle markers ($\circ$), BRSS-EL with $m=4$ as dotted lines with triangle markers ($\triangle$), BRSS-KER with $m=2$ as dashed lines with plus markers ($+$), and BRSS-KER with $m=4$ as dash-dot lines with cross markers ($\times$).}
	    \label{fig:lognormal.cp.full}
	\end{figure}
	
	\begin{figure}[!ht]
	    \centerline{\includegraphics[width=0.65\textwidth]{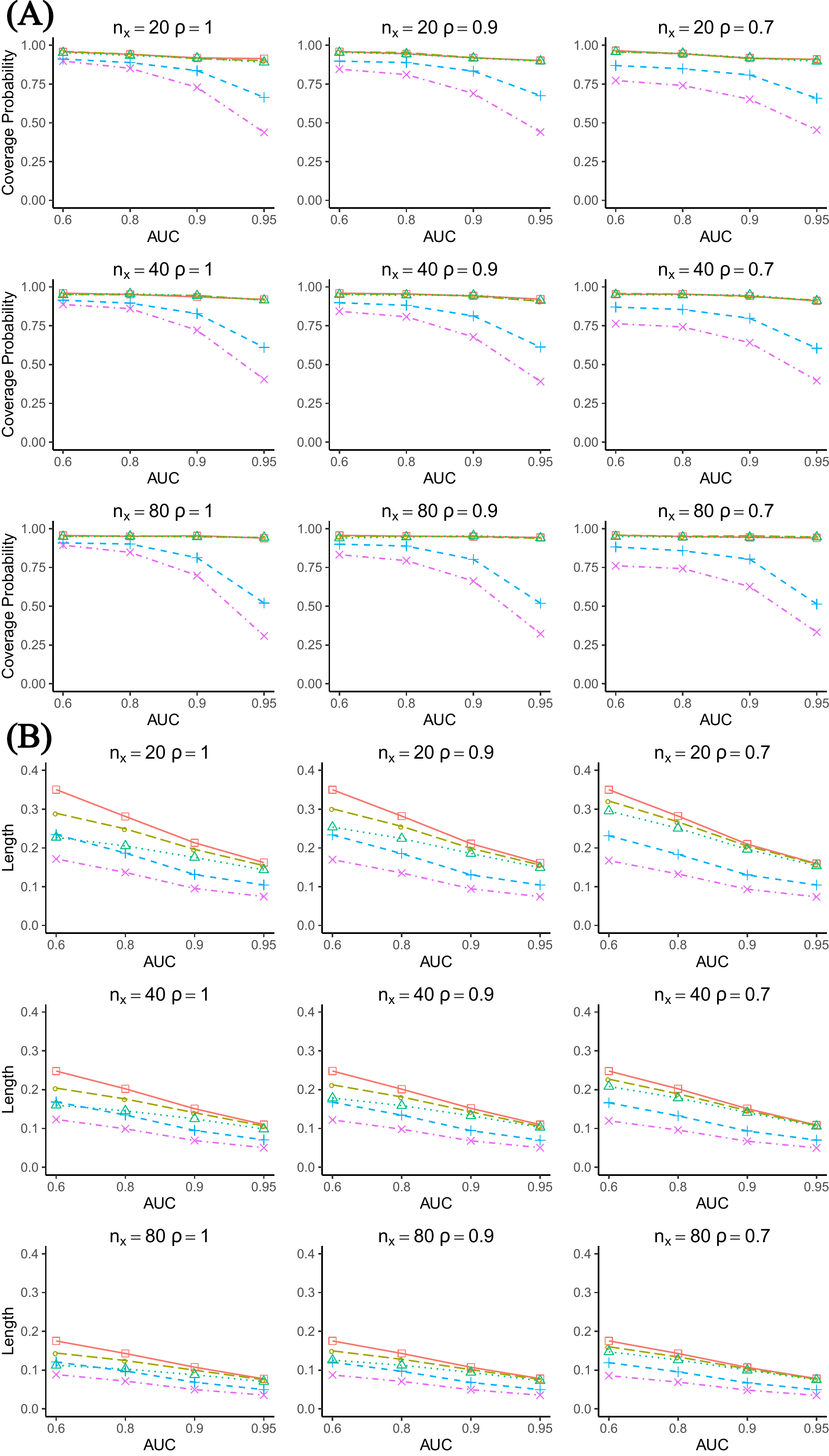}}
	    \caption{Estimated 95\% confidence intervals of the AUC for uniform distributions: (A) coverage probabilities, (B) average lengths. Five lines are shown: SRS-EL as solid lines with square markers ($\square$), BRSS-EL with $m=2$ as long-dashed lines with circle markers ($\circ$), BRSS-EL with $m=4$ as dotted lines with triangle markers ($\triangle$), BRSS-KER with $m=2$ as dashed lines with plus markers ($+$), and BRSS-KER with $m=4$ as dash-dot lines with cross markers ($\times$).}
	    \label{fig:uniform.cp.full}
	\end{figure}

		\begin{figure}[!ht]
	    \centerline{\includegraphics[width=0.6\textwidth]{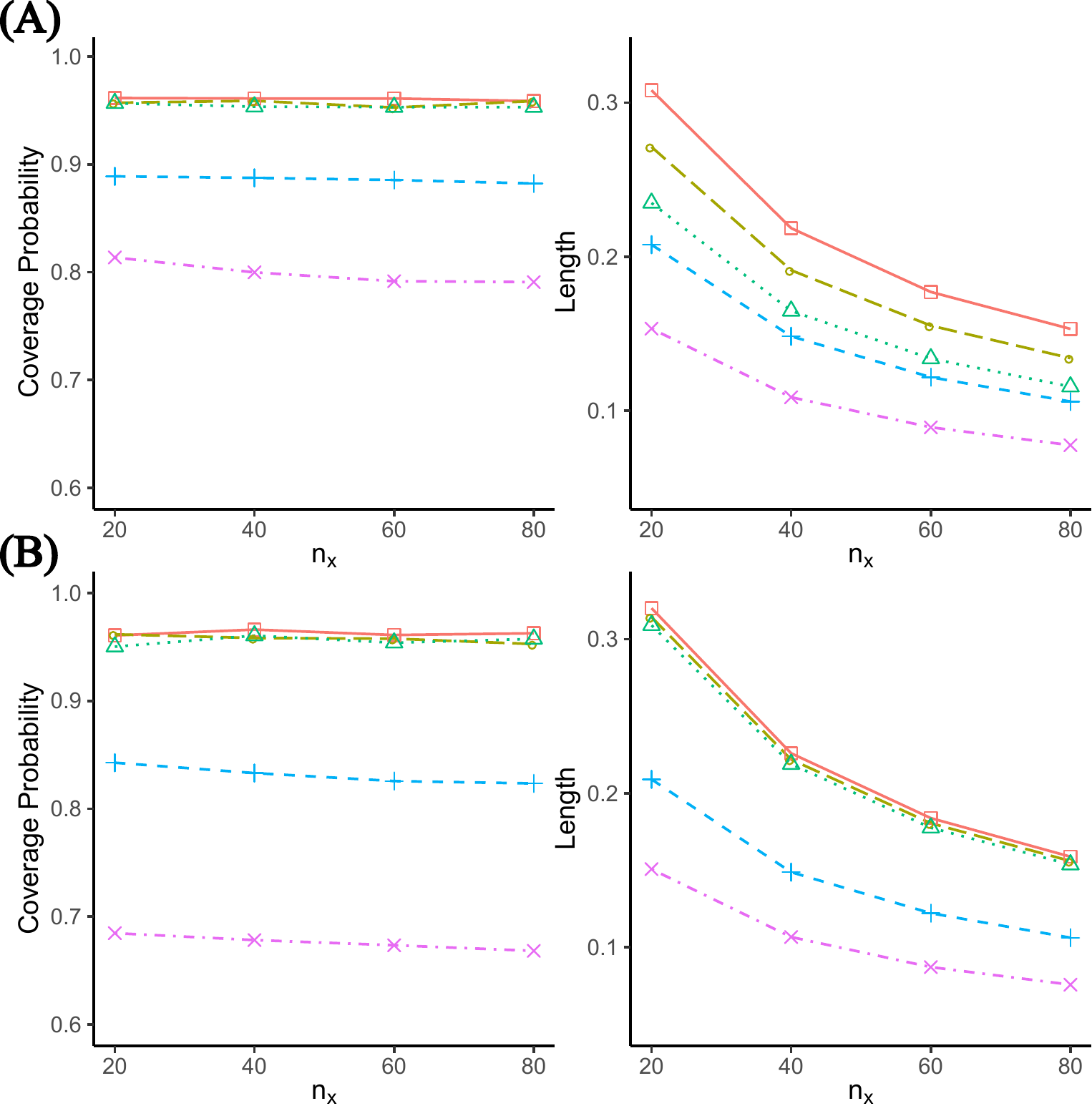}}
	    \caption{Coverage probabilities and average lengths for 95\% confidence intervals of the AUC for the diabetes data: (A) weight is used as concomitant variable, (B) TotChol is used as concomitant variable. Five lines are shown: SRS-EL as solid lines with square markers ($\square$), BRSS-EL with $m=2$ as long-dashed lines with circle markers ($\circ$), BRSS-EL with $m=4$ as dotted lines with triangle markers ($\triangle$), BRSS-KER with $m=2$ as dashed lines with plus markers ($+$), and BRSS-KER with $m=4$ as dash-dot lines with cross markers ($\times$).}
	    \label{fig:diabetes.full}
	\end{figure}
	
		\begin{figure}[!ht]
	    \centering
	    \includegraphics[width=0.6\textwidth]{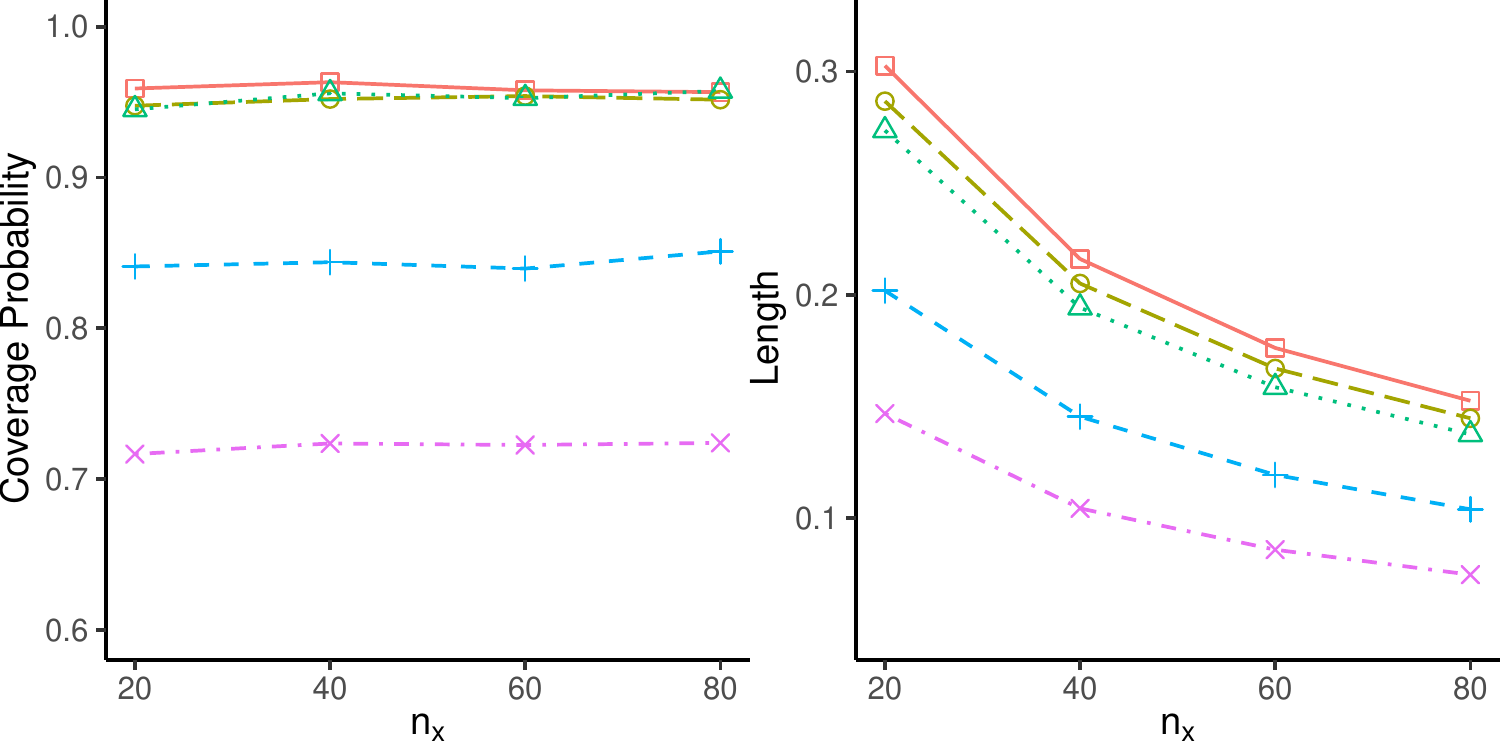}
	    \caption{Coverage probabilities and average lengths for 95\% confidence intervals of the AUC for the CKD data when age is used as the concomitant variable. Five lines are shown: SRS-EL as solid lines with square markers ($\square$), BRSS-EL with $m=2$ as long-dashed lines with circle markers ($\circ$), BRSS-EL with $m=4$ as dotted lines with triangle markers ($\triangle$), BRSS-KER with $m=2$ as dashed lines with plus markers ($+$), and BRSS-KER with $m=4$ as dash-dot lines with cross markers ($\times$).}
	    \label{fig:ckd.full}
	\end{figure}

\clearpage
	
\newpage

\bibliography{References_PST}

\clearpage

\end{document}